\documentclass[prd,preprintnumbers,twocolumn,amsmath,nofootinbib,amssymb]{revtex4}
\usepackage{graphicx,color,dcolumn,booktabs,bm}
\usepackage{longtable,lscape}
\usepackage{makecell}
\usepackage{txfonts}
\usepackage{overpic}
\usepackage{amssymb}
\usepackage{epstopdf}
\usepackage{indentfirst}
\usepackage{feynmf}   
\usepackage{slashed}  
\usepackage{cases}
\usepackage{color}
\usepackage{float}
\usepackage{multirow}
\usepackage{ulem}
\usepackage{enumerate}
\usepackage{graphicx,color,dcolumn,booktabs,bm}
\usepackage{epsfig,dsfont,amssymb,amsmath,amsfonts,amsbsy,mathrsfs}

\graphicspath{{Figures/}} %

\usepackage{hyperref}
\hypersetup{colorlinks,citecolor=blue,anchorcolor=red,menucolor=red, linkcolor=red,filecolor=red,runcolor=red,urlcolor=blue,frenchlinks=true}

\makeatletter
\@addtoreset{equation}{section}
\makeatother

\allowdisplaybreaks

\begin{document}

\title{Unveiling the Composition of the Single-Charm Molecular Pentaquarks: Insights from Radiative Decay and Magnetic Moment}

\author{Fu-Lai Wang$^{1,2,3,5}$}
\email{wangfl2016@lzu.edu.cn}
\author{Si-Qiang Luo$^{1,2,3,4,5}$}
\email{luosq15@lzu.edu.cn}
\author{Xiang Liu$^{1,2,3,4,5}$}
\email{xiangliu@lzu.edu.cn}
\affiliation{$^1$School of Physical Science and Technology, Lanzhou University, Lanzhou 730000, China\\
$^2$Lanzhou Center for Theoretical Physics, Key Laboratory of Theoretical Physics of Gansu Province, Lanzhou University, Lanzhou 730000, China\\
$^3$Key Laboratory of Quantum Theory and Applications of MoE, Lanzhou University,
Lanzhou 730000, China\\
$^4$MoE Frontiers Science Center for Rare Isotopes, Lanzhou University, Lanzhou 730000, China\\
$^5$Research Center for Hadron and CSR Physics, Lanzhou University and Institute of Modern Physics of CAS, Lanzhou 730000, China}

\begin{abstract}
In order to unravel the composition of the isoscalar $DN$, $D^*N$, $D_1N$, and $D_2^*N$ molecular pentaquarks, we carry out a systematic investigation of their M1 and E1 radiative decays and magnetic moment properties. Using the constituent quark model and taking into account the $S$-$D$ wave mixing effect and the coupled channel effect, our analysis yields numerical results indicating that the M1 and E1 radiative decays and the magnetic moment properties of these molecular pentaquarks provide some insights into their inner structures. These results can provide crucial clues for distinguishing their spin-parity quantum numbers and configurations in experimental studies. In addition, we highlight the importance of the electromagnetic properties as key observables for elucidating the inner structures of the observed $\Lambda_c^{+} (2940)$ and $\Lambda_c^{+} (2910)$. We hope that these findings can inspire our experimental colleagues to further explore the family of the single-charm molecular pentaquarks and to investigate the inner structures of the $\Lambda_c^{+} (2940)$ and $\Lambda_c^{+} (2910)$ in future studies.
\end{abstract}
\maketitle

\section{Introduction}\label{sec1}

In the past decades, how to understand the non-perturbative behavior of the strong interaction has received the significant attention from the whole community, especially with the observations of a series of new hadronic states in different experiments \cite{Liu:2013waa,Hosaka:2016pey,Chen:2016qju,Richard:2016eis,Lebed:2016hpi,Brambilla:2019esw,Liu:2019zoy,Chen:2022asf,Olsen:2017bmm,Guo:2017jvc,Meng:2022ozq}. An important reason is that the study of the hadron spectroscopy is one of the most effective approaches to unravelling the non-perturbative dynamics of the strong interaction. Although the hadron family has become more and more abundant with the joint efforts of theorists and experimentalists, the search for exotic states is still an ongoing research issue around the hadron spectroscopy.

Indeed, the observations of more and more new hadronic states close to the corresponding hadronic thresholds have stimulated extensive discussions on the existence of the hadronic molecular states in the past two decades \cite{Liu:2013waa,Hosaka:2016pey,Chen:2016qju,Richard:2016eis,Lebed:2016hpi,Brambilla:2019esw,Liu:2019zoy,Chen:2022asf,Olsen:2017bmm,Guo:2017jvc,Meng:2022ozq}. In particular, the explorations of the hidden-charm molecular pentaquarks have achieved the significant breakthroughs \cite{Aaij:2015tga,Aaij:2019vzc,LHCb:2020jpq,LHCb:2022ogu,Li:2014gra,Karliner:2015ina,Wu:2010jy,Wang:2011rga,Yang:2011wz,Wu:2012md,Chen:2015loa}. As the important experimental progress, the LHCb Collaboration observed three resonance structures, namely the $P_{\psi}^{N}(4312)$, $P_{\psi}^{N}(4440)$, and $P_{\psi}^{N}(4457)$ states, when analyzing the $J/\psi p$ invariant mass spectrum in 2019 \cite{Aaij:2019vzc}. This experimental observation provides the direct evidence to
support the existence of the $\Sigma_c \bar D^{(*)}$ molecular pentaquarks \cite{Li:2014gra,Karliner:2015ina,Wu:2010jy,Wang:2011rga,Yang:2011wz,Wu:2012md,Chen:2015loa}.
Given the current research status of the hidden-charm molecular pentaquarks, it is reasonable to speculate about the existence of the single-charm molecular pentaquark candidates in the hadron spectroscopy.

In 2006, the $\Lambda_c^{+} (2940)$ was first observed in the $D^0p$ invariant mass distributions by the BaBar Collaboration \cite{BaBar:2006itc}. It was later confirmed in the $\Sigma_c(2455) \pi$ decay channel by the Belle Collaboration \cite{Belle:2006xni} and the $D^0p$ channel by LHCb \cite{LHCb:2017jym}. The most likely spin-parity quantum number assignment for the $\Lambda_c^{+} (2940)$ is $J^P=3/2^-$, but the other spin-parity quantum numbers cannot be excluded, as indicated by LHCb \cite{LHCb:2017jym}.
In 2022, Belle reported the discovery of the $\Lambda_c^{+} (2910)$ in the $\Sigma_c(2455) \pi$ invariant mass spectrum \cite{Belle:2022hnm}. For the observed $\Lambda_c^{+} (2940)$ and $\Lambda_c^{+} (2910)$, their masses are lower than the corresponding theoretical results for the $2P$ states of singly charmed baryon from the quenched quark model calculations \cite{Capstick:1986ter,Ebert:2011kk}. Given the existence of the low-mass puzzle for the $\Lambda_c^{+} (2940)$ and $\Lambda_c^{+} (2910)$, and their masses situated just below the $D^*N$ threshold, the $\Lambda_c^{+} (2940)$ and $\Lambda_c^{+} (2910)$ as the $D^*N$ molecular states have stimulated extensive discussions \cite{He:2006is,Garcia-Recio:2008rjt,Dong:2009tg,Dong:2010xv,He:2010zq,Liang:2011zza,Dong:2011ys,Ortega:2012cx,Zhang:2012jk,Ortega:2013fta,Dong:2014ksa,Ortega:2014eoa,Xie:2015zga,Yang:2015eoa,Zhao:2016zhf,Zhang:2019vqe,Wang:2020dhf,Yan:2022nxp,Xin:2023gkf,Ozdem:2023eyz,
Yan:2023ttx,Yue:2024paz}. It is important to note that this is a way to solve the low mass puzzle of the $\Lambda_c^{+} (2940)$ and $\Lambda_c^{+} (2910)$. In fact, recognizing the importance of the unquenched effect, it is still possible to categorize the $\Lambda_c^{+} (2940)$ and $\Lambda_c^{+} (2910)$ as the $2P$ states of singly charmed baryon \cite{Cheng:2006dk,Chen:2007xf,Ebert:2007nw,Zhong:2007gp,Liu:2009zg,Klempt:2009pi,Chen:2009tm,Cheng:2012fq,Lu:2016ctt,Lu:2018utx,Guo:2019ytq,Luo:2019qkm,Gandhi:2019xfw,Gong:2021jkb,Yu:2022ymb,Azizi:2022dpn,Zhang:2022pxc,Yu:2023bxn,Yang:2023fsc}. Obviously, the $\Lambda_c^{+} (2940)$ and $\Lambda_c^{+} (2910)$ as the $D^*N$ molecular states or the $2P$ states of singly charmed baryon have sparked extensive theoretical discussions. However, their inner structures have not been conclusively determined until now. The question of how to distinguish these different assignments to the $\Lambda_c^{+} (2940)$ and $\Lambda_c^{+} (2910)$ is an intriguing and important research issue.

As the important physical observables to reveal the inner structure of the hadron, in this work we propose to systematically study the electromagnetic properties including the M1 and E1 radiative decay widths and the magnetic moment properties of the isoscalar $DN$, $D^*N$, $D_1N$, and $D_2^*N$ molecular pentaquarks, which will be a main task of the present study. Meanwhile, we also discuss the M1 radiative decay widths and the magnetic moment properties of the $\Lambda_c^{+} (2940)$ and $\Lambda_c^{+} (2910)$ in the $2P$ states of singly charmed baryon  \cite{Cheng:2006dk,Chen:2007xf,Ebert:2007nw,Zhong:2007gp,Liu:2009zg,Klempt:2009pi,Chen:2009tm,Cheng:2012fq,Lu:2016ctt,Lu:2018utx,Guo:2019ytq,Luo:2019qkm,Gandhi:2019xfw,Gong:2021jkb,Yu:2022ymb,Azizi:2022dpn,Zhang:2022pxc,Yu:2023bxn,Yang:2023fsc}. To achieve our goal, we adopt the constituent quark model, which has been widely used to discuss the radiative decay widths and the magnetic moment properties of the hadrons in the past decades \cite{Liu:2003ab,Huang:2004tn,Zhu:2004xa,Haghpayma:2006hu,Wang:2016dzu,Deng:2021gnb,Gao:2021hmv,Zhou:2022gra,Wang:2022tib,Li:2021ryu,Schlumpf:1992vq,Schlumpf:1993rm,Cheng:1997kr,Ha:1998gf,Ramalho:2009gk,Girdhar:2015gsa,Menapara:2022ksj,Mutuk:2021epz,Menapara:2021vug,Menapara:2021dzi,Gandhi:2018lez,Dahiya:2018ahb,Kaur:2016kan,Thakkar:2016sog,Shah:2016vmd,Dhir:2013nka,Sharma:2012jqz,Majethiya:2011ry,Sharma:2010vv,Dhir:2009ax,Simonis:2018rld,Ghalenovi:2014swa,Kumar:2005ei,Rahmani:2020pol,Hazra:2021lpa,Gandhi:2019bju,Majethiya:2009vx,Shah:2016nxi,Shah:2018bnr,Ghalenovi:2018fxh,Wang:2022nqs,Mohan:2022sxm,An:2022qpt,Kakadiya:2022pin,Wu:2022gie,Wang:2023bek,Wang:2023aob,Wang:2023ael,Guo:2023fih,Li:2024wxr,Li:2024jlq,Lai:2024jfe}. In our realistic calculations, both the $S$-$D$ wave mixing effect and the coupled channel effect are taken into account, which is similar to the study of their mass spectrum in Refs. \cite{He:2010zq,Chen:2014mwa}. The current investigation can provide some insights into the properties of the single-charm molecular pentaquark candidates and the inner structures of the observed $\Lambda_c^{+} (2940)$ and $\Lambda_c^{+} (2910)$, which may inspire our experimental colleagues to further explore the family of the single-charm molecular pentaquarks and investigate the inner structures of the $\Lambda_c^{+} (2940)$ and $\Lambda_c^{+} (2910)$ in future studies.

The present paper is organized as the follows. After Introduction, we study the M1 and E1 radiative decay widths of the isoscalar $DN$, $D^*N$, $D_1N$, and $D_2^*N$ molecular pentaquarks in Sec. \ref{sec2}. After that, we further investigate the magnetic moment properties of the isoscalar $DN$, $D^*N$, $D_1N$, and $D_2^*N$ molecular pentaquarks in Sec. \ref{sec3}. Among them, we also discuss the M1 radiative decay widths and the magnetic moment properties of the $\Lambda_c^{+} (2940)$ and $\Lambda_c^{+} (2910)$ in the $2P$ states of singly charmed baryon. This work ends with the discussions and conclusions in Sec. \ref{sec4}.

\section{Radiative decay behaviors}\label{sec2}

Before discussing the electromagnetic properties of the single-charm molecular pentaquarks, let's briefly review the research status of their mass spectrum based on the one-boson-exchange (OBE) model and the chiral effective field theory. In Refs. \cite{He:2006is,He:2010zq}, the authors discussed the $D^*N$ molecular pentaquark candidates using the OBE model, and found that several isoscalar $D^*N$ states can be considered as the single-charm molecular pentaquark candidates. In Ref. \cite{Wang:2020dhf}, the authors investigated the $DN$ and $D^*N$ interactions within the chiral effective field theory, and concluded that the isoscalar $DN$ state with $J^P=1/2^-$ and the isoscalar $D^*N$ states with $J^P=(1/2^-,\,3/2^-)$ can be regarded as the single-charm molecular pentaquark candidates. In Ref. \cite{Chen:2014mwa}, the mass spectrum of the $D_1N/D_2^*N$\footnote{In the present work, $D_1$ and $D_2^*$ represent the $D_1(2420)$ and $D_2^*(2460)$ mesons \cite{ParticleDataGroup:2022pth}, respectively. Among them, $D_1(2420)$ should be the mixture of the $1^{1}P_{1}$ and $1^{3}P_{1}$ states of the charmed meson, i.e., $\left|D_1(2420)\right\rangle=-\sin \theta_{1P}\left|1^{1}P_{1}\right\rangle+\cos \theta_{1P}\left|1^{3}P_{1}\right\rangle$ \cite{Godfrey:1986wj,Matsuki:2010zy,Barnes:2002mu,Song:2015fha}.
Under the heavy quark limit, $\theta_{1P}=-54.7^{\circ}$ can be determined by coupling the orbital angular momentum with the spin angular momentum \cite{Godfrey:1986wj,Matsuki:2010zy,Barnes:2002mu,Song:2015fha}.}-type single-charm molecular pentaquark candidates has also been predicted through the OBE model, and they found that it is easier to form the isoscalar $D_1N$ and $D_2^*N$ molecules compared to the isovector $D_1N$ and $D_2^*N$ systems \cite{Chen:2014mwa}. Thus, in the present work we further study the M1 and E1 radiative decay widths and the magnetic moment properties of the isoscalar $DN$, $D^*N$, $D_1N$, and $D_2^*N$ molecular pentaquarks, which can provide the crucial information to decode their spectroscopic behavior.

In comparison to other electromagnetic properties of the hadrons, the radiative decays of the hadronic states may be easier to observe in experimental studies. In particular, the radiative decays of the hadrons can serve as the important clues for unravelling their inner structures. As emphasised in Refs. \cite{BaBar:2008flx,LHCb:2014jvf,Belle:2011wdj,BESIII:2020nbj}, $R= \mathcal{B}(X(3872) \rightarrow \psi(2S) \gamma)/\mathcal{B}(X(3872) \rightarrow J/\psi \gamma)$ can provide the important information for unveiling the inner structure of the $X(3872)$. Thus, the radiative decays of the isoscalar $DN$, $D^*N$, $D_1N$, and $D_2^*N$ molecular pentaquarks can provide the valuable information in the experimental construction of the family of the single-charm molecular pentaquark states.

\subsection{The constituent quark model}

{Quantum Chromodynamics (QCD) describes the interactions between quarks and gluons. However, quantitatively deducing the properties of hadrons from the QCD Lagrangian remains infeasible. To address this, numerous phenomenological models and approaches have been developed over the past several decades. Among these, the constituent quark model stands out as one of the most effective methods for discussing the electromagnetic properties of hadrons, particularly its ability to quantitatively reproduce experimental data on the magnetic moments of decuplet and octet baryons \cite{Schlumpf:1993rm, Kumar:2005ei, Ramalho:2009gk}. Leveraging insights from studying the magnetic moments of these baryons, the constituent quark model has been extensively employed to investigate the radiative decay widths and magnetic moments of hadronic molecular states \cite{Liu:2003ab,Huang:2004tn,Zhu:2004xa,Wang:2016dzu,Li:2021ryu,Deng:2021gnb,Wang:2022tib,Zhou:2022gra,Gao:2021hmv,Wang:2022nqs,Wang:2023aob,Guo:2023fih,Wang:2023bek,Wang:2023ael,Li:2024wxr,Li:2024jlq,Lai:2024jfe}.} In the present work, we quantitatively study the radiative decay widths of the isoscalar $DN$, $D^*N$, $D_1N$, and $D_2^*N$ molecular pentaquarks by adopting the constituent quark model.

\begin{figure}[htbp]
\centering
 \includegraphics[width=0.45\textwidth]{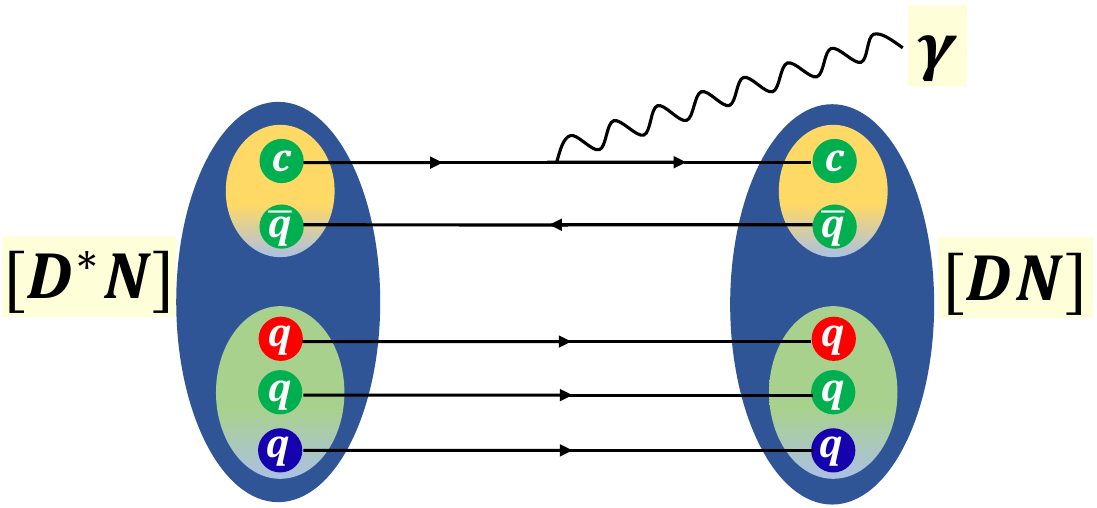}
\caption{Photon emission in the $[D^*N ]\to [DN] \gamma$ process within the constituent quark model. Here, we only present the charm quark in the charmed meson $D^*$ that emit the photon, and the light flavor quarks in the charmed meson $D^*$ and the nucleon $N$ can also emit the photon.}\label{schematicdiagram}
\end{figure}

In the framework of the constituent quark model, the radiative decay of an initial single-charm molecular pentaquark candidate to a final single-charm molecular pentaquark candidate can be described by the emission of the photon from each quark within the initial single-charm molecular pentaquark candidate. In the following, we will give an example to illustrate this point. In Fig. \ref{schematicdiagram}, we present a schematic diagram to illustrate the M1 radiative decay for the $[D^* N] \to [DN] \gamma$ process within the constituent quark model \cite{Ortiz-Pacheco:2023kjn,Ridwan:2024ngc}. In this process, the emission of the photon by the charm quark and the light flavor quark in the charmed meson $D^*$, as well as the light flavor quarks in the nucleon $N$, leads to the $[D^*N] \to [DN] \gamma$ decay.

In the following, we present how to study the radiative decay widths of the isoscalar $DN$, $D^*N$, $D_1N$, and $D_2^*N$ molecular pentaquarks within the constituent quark model. According to Refs. \cite{Dey:1994qi,Simonis:2018rld,Gandhi:2019bju,Hazra:2021lpa,Li:2021ryu,Zhou:2022gra,Wang:2022tib,Rahmani:2020pol,Menapara:2022ksj,Menapara:2021dzi,Gandhi:2018lez,Majethiya:2011ry,Majethiya:2009vx,Shah:2016nxi,Ghalenovi:2018fxh,Wang:2022nqs,Mohan:2022sxm,An:2022qpt,Kakadiya:2022pin,Wang:2023bek,Wang:2023aob,Lai:2024jfe,Wang:2023ael}, the M1 radiative decay width between the hadrons $\Gamma_{H \to H^{\prime}\gamma}^{\rm M1}$ can be linked to the corresponding transition magnetic moment $\mu_{H \to H^{\prime}}$. Thus, we need to define the transition magnetic moment. The transition magnetic moments between the isoscalar $DN$, $D^*N$, $D_1N$, and $D_2^*N$ molecular pentaquarks can be deduced by \cite{Wang:2022nqs,Wang:2023aob,Wang:2023ael,Wang:2023bek,Lai:2024jfe}
\begin{eqnarray}\label{transitionmagneticmoment}
\mu_{H \to H^{\prime}}=\left\langle{J_{H^{\prime}},J_{z}\left|\sum_{j}\hat{\mu}_{zj}^{\rm spin}e^{-i {\bf k}\cdot{\bf r}_j}\right|J_{H},J_{z}}\right\rangle,\label{transitionmagnetic}
\end{eqnarray}
where $J_z={\rm Min}\{J_H,\,J_{H^{\prime}}\}$, $e^{-i{\bf k} \cdot{\bf r}_j}$ is the spatial wave function of the emitted photon, and ${\bf k}$ is the momentum of the emitted photon with $k={(m_{H}^2-m_{H^{\prime}}^2)}/{(2m_{H})}$. In addition, $\hat{\mu}_{zj}^{\rm spin}$ is the spin magnetic moment operator, i.e., \cite{Liu:2003ab,Huang:2004tn,Zhu:2004xa,Haghpayma:2006hu,Wang:2016dzu,Deng:2021gnb,Gao:2021hmv,Zhou:2022gra,Wang:2022tib,Li:2021ryu,Schlumpf:1992vq,Schlumpf:1993rm,Cheng:1997kr,Ha:1998gf,Ramalho:2009gk,Girdhar:2015gsa,Menapara:2022ksj,Mutuk:2021epz,Menapara:2021vug,Menapara:2021dzi,Gandhi:2018lez,Dahiya:2018ahb,Kaur:2016kan,Thakkar:2016sog,Shah:2016vmd,Dhir:2013nka,Sharma:2012jqz,Majethiya:2011ry,Sharma:2010vv,Dhir:2009ax,Simonis:2018rld,Ghalenovi:2014swa,Kumar:2005ei,Rahmani:2020pol,Hazra:2021lpa,Gandhi:2019bju,Majethiya:2009vx,Shah:2016nxi,Shah:2018bnr,Ghalenovi:2018fxh,Wang:2022nqs,Mohan:2022sxm,An:2022qpt,Kakadiya:2022pin,Wu:2022gie,Wang:2023bek,Wang:2023aob,Wang:2023ael,Lai:2024jfe}
\begin{eqnarray}\label{eq:muspin}
\hat{\mu}_{zj}^{\rm spin}&=&\frac{e_j}{2m_j}\hat{\sigma}_{zj}.
\end{eqnarray}
Here, the charge, the mass, and the $z$-component of the Pauli spin operator of the $j$-th constituent of the hadronic state are defined as $e_j$, $m_j$, and $\hat{\sigma}_{zj}$, respectively. Besides the transition magnetic moments, we also can define the amplitude for the electromagnetic transitions \cite{Deng:2016stx}
\begin{equation}\label{electromagnetictransitions}
{\cal A}_{J_{Hz}\to J_{H^{\prime}z}}^E=-i\sqrt{\frac{k}{2}}\left\langle{J_{H^{\prime}},J_{H^{\prime}_z}\left|\sum_{j}{\bf r}_j\cdot {\bm \epsilon}e^{-i {\bf k}\cdot{\bf r}_j}\right|J_{H},J_{H_z}}\right\rangle,
\end{equation}
where ${\bm \epsilon}$ is the polarization vector of the photon.

In addition, the spatial wave function of the emitted photon $e^{-i{\bf k}\cdot{\bf r}_j}$ must be expressed as the following way \cite{Khersonskii:1988krb}
\begin{eqnarray}\label{eq:expikr}
e^{-i{\bf k}\cdot{\bf r}}=\sum\limits_{l=1}^\infty\sum\limits_{m=-l+1}^{l-1}4\pi(-i)^{l-1}j_{l-1}(kr)Y_{l-1m}^*(\Omega_{\bf k})Y_{l-1m}(\Omega_{{\bf r}}),
\end{eqnarray}
where $j_{l-1}(x)$ is the spherical Bessel function, and $Y_{l-1 m}(\Omega_{\bf x})$ is the spherical harmonic function. When we take the lowest order of the right hand of Eq.~(\ref{eq:expikr}), i.e., $l=1$, the obtained results are related to the M1 and E1 radiative decays. Thus, we can further give the general expressions of the corresponding M1 and E1 radiative decay widths as follows \cite{Wang:2022tib,Zhou:2022gra,Wang:2022nqs,Wang:2023aob,Wang:2023ael,Wang:2023bek,Lai:2024jfe,Deng:2016stx}
\begin{eqnarray}\label{eq:M1}
\Gamma_{H \to H^{\prime}\gamma}^{\rm M1}&=&\frac{\alpha_{\rm {EM}}}{2J_{H}+1} \frac{k^{3}}{m_{p}^{2}}\frac{\sum\limits_{J_{H^{\prime}z},J_{Hz}}\left(\begin{array}{ccc} J_{H^{\prime}}&1&J_{H}\\-J_{H^{\prime}z}&0&J_{Hz}\end{array}\right)^2}{\left(\begin{array}{ccc} J_{H^{\prime}}&1&J_{H}\\-J_{z}&0&J_{z}\end{array}\right)^2}\nonumber\\
&&\times\frac{\left|\mu_{H \to H^{\prime}}(l=1)\right|^2}{\mu_N^2},\\
\Gamma_{H \to H^{\prime}\gamma}^{\rm E1}&=&\frac{8\alpha_{\rm EM}k^2}{2J_H+1}\sum_{J_{H^{\prime}z},J_{Hz}}|{\cal A}_{J_{Hz}\to J_{H^{\prime}z}}^E(l=1)|^2.
\end{eqnarray}
Here, $\alpha_{\rm {EM}}$ is the electromagnetic fine structure constant with $\alpha_{\rm {EM}} \approx {1}/{137}$, $m_p$ is the mass of the proton, $\mu_N$ is the nuclear magneton with $\mu_N=\frac{e}{2m_p}$, and the 3-$j$ coefficient is defined by the notation $\left(\begin{array}{ccc} a&b&c\\d&e&f\end{array}\right)$.

For the isoscalar $DN$, $D^*N$, $D_1N$, and $D_2^*N$ molecular pentaquarks, the numerical spatial wave functions will be adopted in our concrete calculations, which can be extracted by solving the coupled channel Schr\"{o}dinger equation based on the OBE effective potentials \cite{Luo:2022cun,Chen:2014mwa}. For the nucleon and the charmed mesons, the simple harmonic oscillator wave function is used to describe their spatial wave functions \cite{Wang:2022tib,Zhou:2022gra,Wang:2022nqs,Wang:2023aob,Wang:2023ael,Wang:2023bek,Lai:2024jfe}, i.e.,
\begin{eqnarray}
&&\phi_{n,l,m}(\beta,{\bf r})\nonumber\\
&&=\sqrt{\frac{2n!}{\Gamma(n+l+\frac{3}{2})}}L_{n}^{l+\frac{1}{2}}(\beta^2r^2)\beta^{l+\frac{3}{2}}{\mathrm e}^{-\frac{\beta^2r^2}{2}}r^l Y_{l m}(\Omega_{\bf r}).
\end{eqnarray}
Here,  the radial, the orbital, and the magnetic quantum numbers of the  hadronic states are defined as $n$, $l$, and $m$, respectively. $L_{n}^{l+\frac{1}{2}}(x)$ is the associated Laguerre polynomial. In our realistic calculations, we need a series of oscillating parameters $\beta$, {which can be extracted from their mass spectra}, and the values of $\beta_{D}=0.601~{\rm GeV}$, $\beta_{D^{*}}=0.516~{\rm GeV}$, $\beta_{c \bar q|1^1P_{1}\rangle}=0.475~{\rm GeV}$, $\beta_{c \bar q|1^3P_{1}\rangle}=0.482~{\rm GeV}$, and $\beta_{c \bar q|1^3P_{2}\rangle}=0.437~{\rm GeV}$ are employed \cite{Godfrey:2015dva}.

In this work, we also discuss the magnetic moment properties of the isoscalar $DN$, $D^*N$, $D_1N$, and $D_2^*N$ molecular pentaquarks. When adopting the constituent quark model, the magnetic moment is calculated by~\cite{Liu:2003ab,Huang:2004tn,Zhu:2004xa,Haghpayma:2006hu,Wang:2016dzu,Deng:2021gnb,Gao:2021hmv,Zhou:2022gra,Wang:2022tib,Li:2021ryu,Schlumpf:1992vq,Schlumpf:1993rm,Cheng:1997kr,Ha:1998gf,Ramalho:2009gk,Girdhar:2015gsa,Menapara:2022ksj,Mutuk:2021epz,Menapara:2021vug,Menapara:2021dzi,Gandhi:2018lez,Dahiya:2018ahb,Kaur:2016kan,Thakkar:2016sog,Shah:2016vmd,Dhir:2013nka,Sharma:2012jqz,Majethiya:2011ry,Sharma:2010vv,Dhir:2009ax,Simonis:2018rld,Ghalenovi:2014swa,Kumar:2005ei,Rahmani:2020pol,Hazra:2021lpa,Gandhi:2019bju,Majethiya:2009vx,Shah:2016nxi,Shah:2018bnr,Ghalenovi:2018fxh,Wang:2022nqs,Mohan:2022sxm,An:2022qpt,Kakadiya:2022pin,Wu:2022gie,Wang:2023bek,Wang:2023aob,Wang:2023ael,Lai:2024jfe}
\begin{eqnarray}
\mu_{H}&=&\left\langle{J_{H},J_{H}\left|\sum_{j}\hat{\mu}_{zj}^{\rm spin}+\hat{\mu}_z^{\rm orbital}\right|J_{H},J_{H}}\right\rangle,
\end{eqnarray}
where $\hat{\mu}_{zj}^{\rm spin}$ has been defined in Eq.~(\ref{eq:muspin}), and $\hat{\mu}_z^{\rm orbital}$ is the orbital magnetic moment, which is given by
\begin{eqnarray}
\hat{\mu}_z^{\rm orbital}&=&\left(\frac{m_{m}}{m_{b}+m_{m}}\frac{e_b}{2m_b}+\frac{m_{b}}{m_{b}+m_{m}}\frac{e_m}{2m_m}\right)\hat{L}_z,
\end{eqnarray}
where the subscripts $b$ and $m$ represent nucleon and charmed mesons, respectively. And we define the $z$-component of the orbital angular momentum operator between the nucleon and the charmed mesons as $\hat{L}_z$.

In the specific studies, we systematically discuss the electromagnetic properties including the M1 and E1 radiative decay widths and the magnetic moment properties of the isoscalar $DN$, $D^*N$, $D_1N$, and $D_2^*N$ molecular pentaquark candidates. In light of the fact that these discussed molecular pentaquark candidates have not yet been definitively identified or observed in experiments, we employ three typical binding energies of $-2$, $-8$, and $-14~{\rm MeV}$ for these molecular pentaquark candidates to quantitatively calculate their M1 and E1 radiative decay widths and magnetic moment properties. This can be attributed to the fact that the hadronic molecular state is a loosely bound state, the reasonable binding energy is around several MeV to several tens MeV only, and there is no significant overlap for the two constituent hadrons in the spatial distribution \cite{Chen:2016qju}\footnote{Given that the mean-square radii of the $D$, $D^*$, $D_1$, $D_2^*$, and $N$ are approximately $0.40\,{\rm fm}$, $0.47\,{\rm fm}$, $0.65\,{\rm fm}$, $0.71\,{\rm fm}$ \cite{Godfrey:2015dva}, and $0.84\,{\rm fm}$ \cite{ParticleDataGroup:2022pth}, respectively, it is reasonable to expect that the mean-square radii of the isoscalar $DN$, $D^*N$, $D_1N$, and $D_2^*N$ molecular pentaquark candidates will be on the order of magnitude of $1\,{\rm fm}$, which can avoid excessive overlap between the corresponding constituent hadrons in their spatial distribution. For example, the corresponding mean-square radii are around $3.16$, $1.77$, and $1.41\,{\rm fm}$ when employing three typical binding energies of $-2$, $-8$, and $-14~{\rm MeV}$ for the isoscalar $DN$ molecular pentaquark candidate.}.
In order to clarify different theoretical explanations for the observed $\Lambda_c^{+} (2940)$ and $\Lambda_c^{+} (2910)$ in the future experiments, we further calculate their M1 radiative decay widths and magnetic moment properties in the frameworks of the $D^*N$ molecular states and the $2P$ states of singly charmed baryon. For the $\Lambda_c^{+} (2940)$ and $\Lambda_c^{+} (2910)$, their masses were measured to be $2939.6^{+1.3}_{-1.5}\,{\rm MeV}$ and $2913.8\pm5.6\pm3.8\,{\rm MeV}$ \cite{ParticleDataGroup:2022pth}, respectively. Here, we need to mention that the experimental uncertainties on the masses of the observed $\Lambda_c^{+} (2940)$ and $\Lambda_c^{+} (2910)$ are relatively small. When calculating the M1 radiative decay widths and the magnetic moment properties of the observed $\Lambda_c^{+} (2940)$ and $\Lambda_c^{+} (2910)$ in the frameworks of the $D^*N$ molecular states and the $2P$ states of singly charmed baryon, we take the central values for the masses of the $\Lambda_c^{+} (2940)$ and $\Lambda_c^{+} (2910)$, i.e., $2939.6\,{\rm MeV}$ and $2913.8\,{\rm MeV}$. Thus, the corresponding binding energies of the $\Lambda_c^{+} (2940)$ and $\Lambda_c^{+} (2910)$ as the $D^*N$ molecular states are $-7.88\,{\rm MeV}$ and $-33.68\,{\rm MeV}$, respectively.
When calculating the radiative decay widths of the isoscalar $DN$, $D^*N$, $D_1N$, and $D_2^*N$ molecular pentaquark candidates, the binding energies of the initial and final molecular pentaquark candidates affect not only the phase spaces of the radiative decay processes, but also the spatial wave functions of the initial and final molecular pentaquark candidates, which result in alterations to the values of the transition magnetic moments $\mu_{H \to H^{\prime}}$ and the amplitudes for the electromagnetic transitions ${\cal A}_{J_{Hz}\to J_{H^{\prime}z}}^E$ associated with the radiative decay processes (see Eqs.~(\ref{transitionmagneticmoment}) and (\ref{electromagnetictransitions})).

\subsection{M1 radiative decays between the isoscalar $DN$ and $D^*N$ molecular pentaquarks}

When calculating the transition magnetic moment between the molecular states within the constituent quark model, the main task is to deduce the matrix element $\left\langle{J_{H^{\prime}},J_{z}\left|\sum\limits_{j}\hat{\mu}_{zj}^{\rm spin}e^{-i {\bf k}\cdot{\bf r}_j}\right|J_{H},J_{z}}\right\rangle$. Thus, we need to construct the wave functions for the initial and final hadronic states, which are composed of the color wave function $\omega^{\rm{color}}$, the flavor and spin wave function $\chi^{\rm{flavor-spin}}$, and the spatial wave function $R^{\rm{spatial}}$. In the concrete calculations, the matrix element $\left\langle{J_{H^{\prime}},J_{z}\left|\sum\limits_{j}\hat{\mu}_{zj}^{\rm spin}e^{-i {\bf k}\cdot{\bf r}_j}\right|J_{H},J_{z}}\right\rangle$ can be written as
\begin{eqnarray}
&&\left\langle{J_{H^{\prime}},J_{z}\left|\sum_{j}\hat{\mu}_{zj}^{\rm spin}e^{-i {\bf k}\cdot{\bf r}_j}\right|J_{H},J_{z}}\right\rangle\nonumber\\
&&=\sum_{j}\left\langle{\chi^{\rm{flavor-spin}}_{H^{\prime}}\left|\hat{\mu}_{zj}^{\rm spin}\right|\chi^{\rm{flavor-spin}}_{H}}\right\rangle\left\langle{R^{\rm{spatial}}_{H^{\prime}}\left|e^{-i {\bf k}\cdot{\bf r}_j}\right|R^{\rm{spatial}}_{H}}\right\rangle.\nonumber\\
\end{eqnarray}
In the above equation, there are no operators which are related to the color wave function. Consequently, the color wave function does not affect the transition magnetic moment between the molecular states, and it is not necessary to consider it explicitly.

To illustrate, we proceed to derive the transition magnetic moment of the $[D^{*}N]({1}/{2}^-)\to [DN]({1}/{2}^-) \gamma$ process as follows.  For the isoscalar $DN$ and $D^*N$ systems, their flavor wave functions can be expressed as $|D^{(*)}N\rangle=\left(|D^{(*)0}p\rangle+|D^{(*)+}n\rangle\right)/\sqrt{2}$. Meanwhile, the spin wave functions $|S, m_{S}\rangle$ of the $DN$ state with $J^P=1/2^-$ and the $D^*N$ state with $J^P=1/2^-$ are
\begin{eqnarray}
\left|\frac{1}{2}, \frac{1}{2}\right\rangle&=&\left|0, 0\right\rangle_D \otimes \left|\frac{1}{2}, \frac{1}{2}\right\rangle_N,\\
\left|\frac{1}{2}, \frac{1}{2}\right\rangle&=&\sqrt{\frac{2}{3}}\left|1, 1\right\rangle_{D^*} \otimes \left|\frac{1}{2}, -\frac{1}{2}\right\rangle_N-\sqrt{\frac{1}{3}}\left|1, 0\right\rangle_{D^*} \otimes \left|\frac{1}{2}, \frac{1}{2}\right\rangle_N,\nonumber\\
\end{eqnarray}
respectively. In addition, the flavor wave functions of the mesons $D^{(*)}$ can be constructed as $D^{(*)0}=c \bar u$ and $D^{(*)+}=c \bar d$, while the spin wave functions $|S, S_3\rangle$ are $\left|0,0\right\rangle=\dfrac{1}{\sqrt{2}}\left(\uparrow\downarrow-\downarrow\uparrow\right)$ and $\left\{
  \begin{array}{l}
    \left|1,1\right\rangle=\uparrow\uparrow\\
    \left|1,0\right\rangle=\dfrac{1}{\sqrt{2}}\left(\uparrow\downarrow+\downarrow\uparrow\right)\\
    \left|1,-1\right\rangle=\downarrow\downarrow
  \end{array}
\right.$ for $D$ and $D^{*}$, respectively. With the above preparation, we can obtain
\begin{eqnarray}
\sum_{j}\left\langle{\chi^{\rm{flavor-spin}}_{[DN]({1}/{2}^-)}\left|\hat{\mu}_{zj}^{\rm spin}\right|\chi^{\rm{flavor-spin}}_{[D^{*}N]({1}/{2}^-)}}\right\rangle=-\frac{\mu_c}{\sqrt{3}}+\frac{\mu_{\overline{u}}+\mu_{\overline{d}}}{2\sqrt{3}}.
\end{eqnarray}
Here, $\mu_{q}={e_q}/{2m_q}$. Furthermore, it is essential to consider the contribution of the factor $\left\langle{R^{\rm{spatial}}_{[DN]({1}/{2}^-)}\left|e^{-i {\bf k}\cdot{\bf r}_j}\right|R^{\rm{spatial}}_{[D^{*}N]({1}/{2}^-)}}\right\rangle$ for the magnetic magnetons of the constituent quarks, which is related to the spatial wave functions of the initial and final states. For details on the calculation of the factor $\left\langle{R^{\rm{spatial}}_{[DN]({1}/{2}^-)}\left|e^{-i {\bf k}\cdot{\bf r}_j}\right|R^{\rm{spatial}}_{[D^{*}N]({1}/{2}^-)}}\right\rangle$, interested readers can refer to Appendix B of Ref. \cite{Wang:2022nqs}.

In this work, the masses of the constituent quarks are $m_u=0.336~{\rm GeV}$, $m_d=0.336~{\rm GeV}$, and $m_c=1.680~{\rm GeV}$ to give the numerical results \cite{Kumar:2005ei}, which was extensively applied to study the radiative decay widths and the magnetic moment properties of the hadrons in the past decades \cite{Li:2021ryu,Zhou:2022gra,Wang:2022tib,Wang:2022nqs,Wang:2023aob,Wang:2023ael,Lai:2024jfe}. In addition, we take the masses of the nucleon and the charmed mesons from the Particle Data Group \cite{ParticleDataGroup:2022pth}.

In order to intuitively assess the reliability of our adopted model, we calculate the transition magnetic moments and the magnetic moments of the nucleons and the ground charmed mesons by adopting the constituent quark model. We then compare our obtained results with other findings to assess the reasonableness of our adopted model. In Table \ref{TMMT}, we collect our obtained transition magnetic moments and magnetic moments of the nucleons and the ground charmed mesons, and compare them with other results \cite{Lai:2024jfe,Wang:2023bek,Simonis:2018rld,Zhou:2022gra}. By comparison, we see that our obtained transition magnetic moments and magnetic moments of the nucleons and the ground charmed mesons are consistent with other results \cite{Lai:2024jfe,Wang:2023bek,Simonis:2018rld,Zhou:2022gra}. In particular, our obtained magnetic moments of the proton and the neutron are extremely close to the corresponding experimental values, with $\mu_p^{\rm Expt} = 2.79\,\mu_N$ and $\mu_n^{\rm Expt} = -1.91\,\mu_N$ \cite{ParticleDataGroup:2022pth}. Thus, our adopted model is reasonable when discussing the transition magnetic moments and the magnetic moments of the nucleons and the ground charmed mesons. This enables us to offer the credible theoretical insights for the experimental investigation of the radiative decay widths and the magnetic moment properties of the isoscalar $DN$, $D^*N$, $D_1N$, and $D_2^*N$ molecular pentaquark candidates.

\renewcommand\tabcolsep{0.23cm}
\renewcommand{\arraystretch}{1.50}
\begin{table}[!htbp]
\centering
\caption{The transition magnetic moments and the magnetic moments of the nucleons and the ground charmed mesons, and compared them with other results. Here, the transition magnetic moments and the magnetic moments of the hadrons are in units of $\mu_{N}$.}\label{TMMT}
\begin{tabular}{c|c|c|c}
\toprule[1.0pt]
\multirow{3}{*}{$\mu_{H \to H^{\prime}}$}&Decays&\multicolumn{1}{c|}{Our result}  &  \multicolumn{1}{c}{Other results} \\\cline{2-4}			
&$D^{*0} \to D^{0} \gamma$ & $2.17$ &$2.17$ \cite{Lai:2024jfe},\,$2.13$ \cite{Wang:2023bek}\\
&$D^{*+} \to D^{+} \gamma$ & $-0.54$ &$-0.54$ \cite{Lai:2024jfe},\,$-0.54$ \cite{Simonis:2018rld}\\\midrule[1.0pt]
\multirow{5}{*}{$\mu_{H}$}&Hadrons &  \multicolumn{1}{c|}{Our result}  &  \multicolumn{1}{c}{Other results} \\\cline{2-4}		
&$p$ & $2.79$ & $2.79$ \cite{ParticleDataGroup:2022pth}\\ 			
&$n$&$-1.86$&$-1.91$ \cite{ParticleDataGroup:2022pth}\\	
&$D^{*0}$ & $-1.49$ &$-1.49$ \cite{Lai:2024jfe},\,$-1.49$ \cite{Zhou:2022gra}\\
&$D^{*+}$ & $1.30$ &$1.30$ \cite{Lai:2024jfe},\,$1.30$ \cite{Zhou:2022gra}\\
\bottomrule[1.0pt]
\end{tabular}
\end{table}

In the following, we proceed to study the M1 radiative decay widths between the isoscalar $DN$ and $D^*N$ molecular pentaquarks. In Table \ref{DecayDNDstarN}, and the numerical results of the transition magnetic moments and the M1 radiative decay widths between the isoscalar $DN$ and $D^*N$ molecules are presented through the single channel analysis. As shown in Table \ref{DecayDNDstarN}, the transition magnetic moments of the $[D^*N]({3}/{2}^-) \to [DN]({1}/{2}^-) \gamma$ and $[D^*N]({1}/{2}^-) \to [DN]({1}/{2}^-) \gamma$ processes are approximately $0.63 \sim 0.66\,\mu_N$ and $-0.44 \sim -0.43\,\mu_N$, respectively. Furthermore, the M1 radiative decay widths of the $[D^*N]({3}/{2}^-) \to [DN]({1}/{2}^-) \gamma$ and $[D^*N]({1}/{2}^-) \to [DN]({1}/{2}^-) \gamma$ processes are about $4.36 \sim 4.71\,{\rm keV}$ and $4.03 \sim 4.15\,{\rm keV}$, respectively. For the $[D^*N]({1}/{2}^-) \to [D^*N]({3}/{2}^-) \gamma$ process, the transition magnetic moment is around $-0.47 \sim -0.43\,\mu_N$, and the corresponding M1 radiative decay width is zero when taking the same binding energies for the initial and final molecular pentaquarks. This phenomenon is caused by the phase space being zero. When the binding energies of the isoscalar $D^*N$ state with $J^P=1/2^-$ and the isoscalar $D^*N$ state with $J^P=3/2^-$ are varied between $-2$ and $-14~{\rm MeV}$, the M1 radiative decay width for the $[D^*N]({1}/{2}^-) \to [D^*N]({3}/{2}^-) \gamma$ process is less than $0.003\,{\rm keV}$, this is mainly due to the limited phase space.

\renewcommand\tabcolsep{0.13cm}
\renewcommand{\arraystretch}{1.50}
\begin{table}[!htbp]
\centering
\caption{The transition magnetic moments and the M1 radiative decay widths between the isoscalar $DN$ and $D^*N$ molecules. Given that these discussed molecular pentaquark candidates have not yet been definitively identified or observed in experiments, we utilize three typical binding energies of $-2$, $-8$, and $-14~{\rm MeV}$ for these molecular pentaquark candidates to quantitatively calculate their M1 radiative decay widths.}\label{DecayDNDstarN}
\begin{tabular}{c|c|c}\toprule[1pt]
Radiative decays&$\mu_{H\to H^{\prime}}\,(\mu_{N})$&$\Gamma_{H \to H^{\prime}\gamma}^{\rm M1}\,({\rm keV})$\\\midrule[1.0pt]
\multicolumn{3}{c}{Single channel analysis}\\
\cline{1-3}
$[D^*N](\frac{3}{2}^-) \to [DN](\frac{1}{2}^-) \gamma$&$0.63$,\,$0.66$,\,$0.66$&$4.36$,\,$4.67$,\,$4.71$\\
$[D^*N](\frac{1}{2}^-) \to [DN](\frac{1}{2}^-) \gamma$&$-0.44$,\,$-0.44$,\,$-0.43$&$4.14$,\,$4.15$,\,$4.03$\\
$[D^*N](\frac{1}{2}^-) \to [D^*N](\frac{3}{2}^-) \gamma$&$-0.47$,\,$-0.45$,\,$-0.43$&0.003\protect\footnote{In this work, the maximum value of the M1 radiative decay width is listed for the decay process involving the same initial and final system, which is obtained by varying the binding energies of the initial and final molecules.}\\\midrule[1.0pt]
\multicolumn{3}{c}{$S$-$D$ wave mixing analysis}\\
\cline{1-3}
$[D^*N](\frac{3}{2}^-) \to [DN](\frac{1}{2}^-) \gamma$&$0.62$,\,$0.63$,\,$0.63$&$4.15$,\,$4.30$,\,$4.26$\\
$[D^*N](\frac{1}{2}^-) \to [DN](\frac{1}{2}^-) \gamma$&$-0.42$,\,$-0.41$,\,$-0.40$&$3.84$,\,$3.71$,\,$3.54$\\
$[D^*N](\frac{1}{2}^-) \to [D^*N](\frac{3}{2}^-) \gamma$&$-0.47$,\,$-0.45$,\,$-0.45$&0.003\\\midrule[1.0pt]
\multicolumn{3}{c}{Coupled channel analysis}\\
\cline{1-3}
$[D^*N](\frac{3}{2}^-) \to [DN](\frac{1}{2}^-) \gamma$&$0.65$,\,$0.69$,\,$0.70$&$4.68$,\,$5.12$,\,$5.25$\\
$[D^*N](\frac{1}{2}^-) \to [DN](\frac{1}{2}^-) \gamma$&$-0.41$,\,$-0.40$,\,$-0.39$&$3.68$,\,$3.44$,\,$3.23$\\
\bottomrule[1pt]
\end{tabular}
\end{table}

Similar to the study of the mass spectrum of the isoscalar $DN$ and $D^*N$ molecular pentaquarks in Ref. \cite{He:2010zq}, we can further consider the contributions of the $S$-$D$ wave mixing effect and the coupled channel effect to discuss their M1 radiative decay widths. Considering the contribution of the $S$-$D$ wave mixing effect, the spin-orbit wave functions $|^{2S+1}L_J\rangle$ for the isoscalar $DN$ state with $J^P=1/2^-$, the isoscalar $D^*N$ state with $J^P=1/2^-$, and the isoscalar $D^*N$ state with $J^P=3/2^-$ are $|{}^2\mathbb{S}_{{1}/{2}}\rangle$, $|{}^2\mathbb{S}_{{1}/{2}}\rangle/|{}^4\mathbb{D}_{{1}/{2}}\rangle$, and $|{}^4\mathbb{S}_{{3}/{2}}\rangle/|{}^2\mathbb{D}_{{3}/{2}}\rangle/|{}^4\mathbb{D}_{{3}/{2}}\rangle$ \cite{He:2010zq}, respectively. Here, $S$, $L$, and $J$ are the spin, the orbital angular momentum, and the total angular momentum of the corresponding channel. For the isoscalar $DN$ state with $J^P=1/2^-$, we can further take into account the influence of the coupled channel effect by including the $DN$ and $D^*N$ channels. When discussing the transition magnetic moment of the $D$-wave channel, it is necessary to expand the spin-orbital wave function $|{ }^{2 S+1} L_{J}\rangle$ by the coupling of the orbital wave function $Y_{L m_{L}}$ and the spin wave function $\left|S, m_{S}\right\rangle$,  i.e., $\left|{ }^{2 S+1} L_{J}\right\rangle=\sum_{m_{L},m_{S}} C_{L m_{L, S m_{S}}}^{J,M}Y_{L m_{L}} \left|S, m_{S}\right\rangle$ \cite{Wang:2022tib,Zhou:2022gra,Wang:2022nqs,Wang:2023aob,Wang:2023ael,Wang:2023bek,Lai:2024jfe}.

In Table \ref{DecayDNDstarN}, we present our obtained transition magnetic moments and M1 radiative decay widths between the isoscalar $DN$ and $D^*N$ molecular pentaquarks through the $S$-$D$ wave mixing analysis and the coupled channel analysis, respectively. By comparing the numerical results obtained from the single channel analysis, we find that the $S$-$D$ wave mixing effect and the coupled channel effect do not significantly influence the transition magnetic moments and the M1 radiative decay widths between the isoscalar $DN$ and $D^*N$ molecular pentaquarks, and the change of their M1 radiative decay widths is less than $1~{\rm keV}$.

To compare the M1 radiative decays between the $[D^*N] \to [DN] \gamma$ and $D^* \to D \gamma$ processes, we calculate the M1 radiative decay width of the $D^* \to D \gamma$ process by adopting the constituent quark model. The obtained numerical results are presented in the following:
\begin{eqnarray}
\renewcommand\tabcolsep{1.50cm}
\renewcommand{\arraystretch}{1.50}
\begin{array}{*{2}c}
\hline
\Gamma_{D^{*+} \to D^{+} \gamma}\,(\mathrm{keV})~~~~~~~~~~~~&2.00  \nonumber\\
\Gamma_{D^{*0} \to D^{0} \gamma}\,(\mathrm{keV})~~~~~~~~~~~~&33.53  \\
\hline
\end{array}.
\end{eqnarray}
Here, we need to mention that our obtained M1 radiative decay width of the $D^{*+} \to D^{+} \gamma$ process is close to the corresponding experimental value $1.33\,\rm{keV}$ \cite{ParticleDataGroup:2022pth}. By comparing the M1 radiative decays of the $[D^*N] \to [DN] \gamma$ and $D^* \to D \gamma$ processes, our findings indicate that the M1 radiative decay widths of the $[D^*N] \to [DN] \gamma$ processes fall within the M1 radiative decay widths of the $D^{*+} \to D^{+} \gamma$ and $D^{*0} \to D^{0} \gamma$ processes, but the M1 radiative decay widths of the $[D^*N] \to [DN] \gamma$ processes are smaller than the average value of the M1 radiative decay widths of the $D^{*+} \to D^{+} \gamma$ and $D^{*0} \to D^{0} \gamma$ processes. Therefore, the spectator particle nucleon plays an important role when discussing the M1 radiative decays of the $[D^*N] \to [DN] \gamma$ processes. In fact, the conclusion that the spectator particle plays a role has been obtained through the calculation of the decay widths of the $D D \pi$ and $D D \gamma$ channels for the $T_{cc}(3875)^+$ under the $DD^*$ molecular picture \cite{Meng:2021jnw,Ling:2021bir}.

\subsection{M1 radiative decays of the $\Lambda_c^{+} (2940)$ and $\Lambda_c^{+} (2910)$ under the molecular state and the charmed baryon interpretations}

Since the discoveries of the $\Lambda_c^{+}(2940)$ and $\Lambda_c^{+}(2910)$, they have sparked extensive theoretical discussions \cite{He:2006is,Garcia-Recio:2008rjt,Dong:2009tg,Dong:2010xv,He:2010zq,Liang:2011zza,Dong:2011ys,Ortega:2012cx,Zhang:2012jk,Ortega:2013fta,Dong:2014ksa,Ortega:2014eoa,Xie:2015zga,Yang:2015eoa,Zhao:2016zhf,Zhang:2019vqe,Wang:2020dhf,Yan:2022nxp,Xin:2023gkf,Ozdem:2023eyz,Yan:2023ttx,
Yue:2024paz,Cheng:2006dk,Chen:2007xf,Ebert:2007nw,Zhong:2007gp,Liu:2009zg,Klempt:2009pi,Chen:2009tm,Cheng:2012fq,Lu:2016ctt,Lu:2018utx,Guo:2019ytq,Luo:2019qkm,Gandhi:2019xfw,Gong:2021jkb,Yu:2022ymb,Azizi:2022dpn,Zhang:2022pxc,Yu:2023bxn,Yang:2023fsc}. However, their inner structures and spin-parity quantum numbers have not been conclusively determined until now. In the following, we discuss whether the spin-parity quantum numbers of the $\Lambda_c^{+} (2940)$ and $\Lambda_c^{+} (2910)$ in the framework of the $D^*N$ molecular states \cite{He:2006is,Garcia-Recio:2008rjt,Dong:2009tg,Dong:2010xv,He:2010zq,Liang:2011zza,Dong:2011ys,Ortega:2012cx,Zhang:2012jk,Ortega:2013fta,Dong:2014ksa,Ortega:2014eoa,Xie:2015zga,Yang:2015eoa,Zhao:2016zhf,Zhang:2019vqe,Wang:2020dhf,Yan:2022nxp,Xin:2023gkf,Ozdem:2023eyz,Yan:2023ttx,Yue:2024paz} can be determined by studying the related M1 radiative decay widths.
\begin{itemize}
  \item If the $\Lambda_c^{+} (2940)$ and $\Lambda_c^{+} (2910)$ can be regarded as the $D^*N$ molecular states with $I(J^P)=0(3/2^-)$ and $0(1/2^-)$, we find
\begin{eqnarray}
\Gamma_{\Lambda_c^{+} (2940) \to [DN]({1}/{2}^-) \gamma}^{\rm M1}&=&4.68\sim5.13\,{\rm keV},\\ \Gamma_{\Lambda_c^{+} (2910) \to [DN]({1}/{2}^-) \gamma}^{\rm M1}&=&1.50\sim1.63\,{\rm keV},\\
\frac{\Gamma_{\Lambda_c^{+} (2940) \to [DN]({1}/{2}^-) \gamma}^{\rm M1}}{\Gamma_{\Lambda_c^{+} (2910) \to [DN]({1}/{2}^-) \gamma}^{\rm M1}}&=&2.88\sim3.42.
\end{eqnarray}
  \item If the $\Lambda_c^{+} (2940)$ and $\Lambda_c^{+} (2910)$ can be interpreted as the $D^*N$ molecular states with $I(J^P)=0(1/2^-)$ and $0(3/2^-)$, we obtain
\begin{eqnarray}
\Gamma_{\Lambda_c^{+} (2940) \to [DN]({1}/{2}^-) \gamma}^{\rm M1}&=&3.46\sim4.17\,{\rm keV},\\ \Gamma_{\Lambda_c^{+} (2910) \to [DN]({1}/{2}^-) \gamma}^{\rm M1}&=&2.05\sim2.61\,{\rm keV},\\
\frac{\Gamma_{\Lambda_c^{+} (2940) \to [DN]({1}/{2}^-) \gamma}^{\rm M1}}{\Gamma_{\Lambda_c^{+} (2910) \to [DN]({1}/{2}^-) \gamma}^{\rm M1}}&=&1.33\sim2.04.
\end{eqnarray}
\end{itemize}
Here, we perform the single channel analysis, the $S$-$D$ wave mixing analysis, and the coupled channel analysis, and take the binding energy of the isoscalar $DN$ state with $J^P=1/2^-$ to be $-8~{\rm MeV}$. Thus, we conclude that the spin-parity quantum numbers of the $\Lambda_c^{+} (2940)$ and $\Lambda_c^{+} (2910)$ in the framework of the $D^*N$ molecular states \cite{He:2006is,Garcia-Recio:2008rjt,Dong:2009tg,Dong:2010xv,He:2010zq,Liang:2011zza,Dong:2011ys,Ortega:2012cx,Zhang:2012jk,Ortega:2013fta,Dong:2014ksa,Ortega:2014eoa,Xie:2015zga,Yang:2015eoa,Zhao:2016zhf,Zhang:2019vqe,Wang:2020dhf,Yan:2022nxp,Xin:2023gkf,Ozdem:2023eyz,Yan:2023ttx,Yue:2024paz} can be determined by studying the related M1 radiative decay widths, especially for the ratio ${\Gamma_{\Lambda_c^{+} (2940) \to [DN]({1}/{2}^-) \gamma}^{\rm M1}}/{\Gamma_{\Lambda_c^{+} (2910) \to [DN]({1}/{2}^-) \gamma}^{\rm M1}}$.

Besides the explanation of the $\Lambda_c^{+} (2940)$ and $\Lambda_c^{+} (2910)$ as the $D^*N$ molecules \cite{He:2006is,Garcia-Recio:2008rjt,Dong:2009tg,Dong:2010xv,He:2010zq,Liang:2011zza,Dong:2011ys,Ortega:2012cx,Zhang:2012jk,Ortega:2013fta,Dong:2014ksa,Ortega:2014eoa,Xie:2015zga,Yang:2015eoa,Zhao:2016zhf,Zhang:2019vqe,Wang:2020dhf,Yan:2022nxp,Xin:2023gkf,Ozdem:2023eyz,Yan:2023ttx,Yue:2024paz}, they can also be interpreted as the singly charmed baryons $\Lambda_c(2P,\,3/2^-)$ and $\Lambda_c(2P,\,1/2^-)$ \cite{Cheng:2006dk,Chen:2007xf,Ebert:2007nw,Zhong:2007gp,Liu:2009zg,Klempt:2009pi,Chen:2009tm,Cheng:2012fq,Lu:2016ctt,Lu:2018utx,Guo:2019ytq,Luo:2019qkm,Gandhi:2019xfw,Gong:2021jkb,Yu:2022ymb,Azizi:2022dpn,Zhang:2022pxc,Yu:2023bxn,Yang:2023fsc}, respectively. In the following, we examine the potential for distinguishing different assignments to the $\Lambda_c^{+} (2940)$ and $\Lambda_c^{+} (2910)$ through the analysis of the related M1 radiative decay widths.
\begin{itemize}
\item If the $\Lambda_c^{+} (2940)$ and $\Lambda_c^{+} (2910)$ can be explained as the $D^*N$ molecular states with $I(J^P)=0(3/2^-)$ and $0(1/2^-)$ \cite{He:2006is,Garcia-Recio:2008rjt,Dong:2009tg,Dong:2010xv,He:2010zq,Liang:2011zza,Dong:2011ys,Ortega:2012cx,Zhang:2012jk,Ortega:2013fta,Dong:2014ksa,Ortega:2014eoa,Xie:2015zga,Yang:2015eoa,Zhao:2016zhf,Zhang:2019vqe,Wang:2020dhf,Yan:2022nxp,Xin:2023gkf,Ozdem:2023eyz,Yan:2023ttx,Yue:2024paz}, the M1 radiative decay width of the $\Lambda_c^{+} (2940) \to \Lambda_c^{+} (2910) \gamma$ process is about $10\,{\rm eV}$.
\item If the $\Lambda_c^{+} (2940)$ and $\Lambda_c^{+} (2910)$ can be assigned as the singly charmed baryons $\Lambda_c(2P,\,3/2^-)$ and $\Lambda_c(2P,\,1/2^-)$ \cite{Cheng:2006dk,Chen:2007xf,Ebert:2007nw,Zhong:2007gp,Liu:2009zg,Klempt:2009pi,Chen:2009tm,Cheng:2012fq,Lu:2016ctt,Lu:2018utx,Guo:2019ytq,Luo:2019qkm,Gandhi:2019xfw,Gong:2021jkb,Yu:2022ymb,Azizi:2022dpn,Zhang:2022pxc,Yu:2023bxn,Yang:2023fsc}, the M1 radiative decay width of the $\Lambda_c^{+} (2940) \to \Lambda_c^{+} (2910) \gamma$ process is around $2\,{\rm eV}$. In the specific calculations, $\beta_{\rho}=0.256~{\rm GeV}$ and $\beta_{\lambda}=0.153~{\rm GeV}$ \cite{Luo:2023sne} are employed in the simple harmonic oscillator wave functions for the $\Lambda_c(2P,\,3/2^-)$ and $\Lambda_c(2P,\,1/2^-)$, while we take the diquark mass to be $m_{[ud]}=0.710~{\rm GeV}$ \cite{Ebert:2010af,Wang:2016dzu,Gao:2021hmv,Guo:2023fih}.
\end{itemize}
Although there exists the obvious difference for the M1 radiative decay width of the $\Lambda_c^{+} (2940) \to \Lambda_c^{+} (2910) \gamma$ process between the hadronic molecular state and the singly charmed baryon interpretations for the $\Lambda_c^{+} (2940)$ and $\Lambda_c^{+} (2910)$, the corresponding M1 radiative decay width is too small. This presents a significant challenge for the future experimental studies. For the $\Lambda_c^{+} (2940) \to \Lambda_c^{+} (2910) \gamma$ process, the relevant phase space values are identical, but the amplitudes differ between the hadronic molecular state and the singly charmed baryon interpretations for the $\Lambda_c^{+} (2940)$ and $\Lambda_c^{+} (2910)$. Here, we need to point out that this discrepancy arises primarily because the hadronic molecular state and the singly charmed baryon interpretations for the $\Lambda_c^{+} (2940)$ and $\Lambda_c^{+} (2910)$ exhibit distinct flavor, spin, and spatial wave functions.

\subsection{M1 radiative decays between the isoscalar $D_1N$ and $D_2^*N$ molecular pentaquarks}

In this subsection, we study the M1 radiative decay widths between the isoscalar $D_1N$ and $D_2^*N$ molecular pentaquarks by conducting the $S$-$D$ wave mixing analysis and the coupled channel analysis \cite{Chen:2014mwa}. When considering the $S$-$D$ wave mixing effect, the spin-orbit wave functions $|^{2S+1}L_J\rangle$ for the isoscalar $D_1N$ state with $J^P=1/2^+$, the isoscalar $D_1N$ state with $J^P=3/2^+$, the isoscalar $D_2^*N$ state with $J^P=3/2^+$, and the isoscalar $D_2^*N$ state with $J^P=5/2^+$ are $|{}^2\mathbb{S}_{{1}/{2}}\rangle/|{}^4\mathbb{D}_{{1}/{2}}\rangle$, $|{}^4\mathbb{S}_{{3}/{2}}\rangle/|{}^2\mathbb{D}_{{3}/{2}}\rangle/|{}^4\mathbb{D}_{{3}/{2}}\rangle$, $|{}^4\mathbb{S}_{{3}/{2}}\rangle/|{}^4\mathbb{D}_{{3}/{2}}\rangle/|{}^6\mathbb{D}_{{3}/{2}}\rangle$, and $|{}^5\mathbb{S}_{{5}/{2}}\rangle/|{}^4\mathbb{D}_{{5}/{2}}\rangle/|{}^6\mathbb{D}_{{5}/{2}}\rangle$ \cite{Chen:2014mwa}, respectively. For the isoscalar $D_1N$ state with $J^P=3/2^+$, we can further consider the influence of the coupled channel effect by including the $D_1N$ and $D_2^*N$ channels \cite{Chen:2014mwa}.

\renewcommand\tabcolsep{0.13cm}
\renewcommand{\arraystretch}{1.50}
\begin{table}[!htbp]
\centering
\caption{The transition magnetic moments and the M1 radiative decay widths between the isoscalar $D_1N$ and $D_2^*N$ molecules. Given that these discussed molecular pentaquark candidates have not yet been definitively identified or observed in experiments, we utilize three typical binding energies of $-2$, $-8$, and $-14~{\rm MeV}$ for these molecular pentaquark candidates to quantitatively calculate their M1 radiative decay widths.}\label{DecayD1ND2starN}
\begin{tabular}{c|c|c}\toprule[1pt]
Radiative decays&$\mu_{H\to H^{\prime}}\,(\mu_{N})$&$\Gamma_{H \to H^{\prime}\gamma}^{\rm M1}\,({\rm keV})$\\\midrule[1.0pt]
\multicolumn{3}{c}{$S$-$D$ wave mixing analysis}\\
\cline{1-3}
$[D_2^*N](\frac{5}{2}^+) \to [D_1N](\frac{3}{2}^+) \gamma$&$0.37$,\,$0.37$,\,$0.37$&$0.05$,\,$0.05$,\,$0.05$\\
$[D_2^*N](\frac{3}{2}^+) \to [D_1N](\frac{3}{2}^+) \gamma$&$-0.19$,\,$-0.19$,\,$-0.20$&$0.01$,\,$0.01$,\,$0.01$\\
$[D_2^*N](\frac{3}{2}^+) \to [D_1N](\frac{1}{2}^+) \gamma$&$0.42$,\,$0.41$,\,$0.41$&$0.04$,\,$0.04$,\,$0.04$\\
$[D_1N](\frac{1}{2}^+) \to [D_1N](\frac{3}{2}^+) \gamma$&$-0.29$,\,$-0.27$,\,$-0.27$&$0.001$\\
$[D_2^*N](\frac{3}{2}^+) \to [D_2^*N](\frac{5}{2}^+) \gamma$&$-0.45$,\,$-0.41$,\,$-0.38$&$0.004$\\\midrule[1.0pt]
\multicolumn{3}{c}{Coupled channel analysis}\\
\cline{1-3}
$[D_2^*N](\frac{5}{2}^+) \to [D_1N](\frac{3}{2}^+) \gamma$&$0.38$,\,$0.38$,\,$0.39$&$0.06$,\,$0.06$,\,$0.06$\\
$[D_2^*N](\frac{3}{2}^+) \to [D_1N](\frac{3}{2}^+) \gamma$&$-0.18$,\,$-0.17$,\,$-0.17$&$0.01$,\,$0.01$,\,$0.01$\\
$[D_1N](\frac{1}{2}^+) \to [D_1N](\frac{3}{2}^+) \gamma$&$-0.27$,\,$-0.24$,\,$-0.22$&$0.001$\\
\bottomrule[1pt]
\end{tabular}
\end{table}

\begin{figure}[htbp]
\centering
 \includegraphics[width=0.49\textwidth]{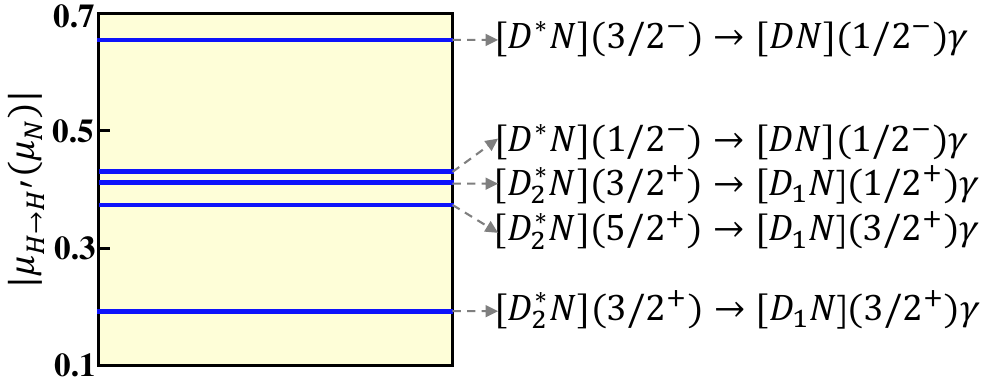}
\caption{The comparison focuses on the absolute values of the transition magnetic moments for the $[D^*N ]\to [DN] \gamma$ and $[D_2^*N] \to [D_1N] \gamma$ processes.}\label{TMMCDHDT}
\end{figure}

In Table \ref{DecayD1ND2starN}, we summary our obtained transition magnetic moments and M1 radiative decay widths between the isoscalar $D_1N$ and $D_2^*N$ molecular pentaquarks. From Table \ref{DecayD1ND2starN}, we summarize several interesting observations:
\begin{enumerate}[(i)]
\item The M1 radiative decay widths between the isoscalar $D_1N$ and $D_2^*N$ molecular pentaquarks is essential for unravelling their inner structures, which can be used to distinguish their spin-parity quantum numbers in the future experiments. For example, there is a significant difference of the M1 radiative decay widths between the $[D_2^*N]({5}/{2}^+) \to [D_1N]({3}/{2}^+) \gamma$ and $[D_2^*N]({3}/{2}^+) \to [D_1N]({3}/{2}^+) \gamma$ processes.
\item By comparing the numerical results obtained from the $S$-$D$ wave mixing analysis, the coupled channel effect is not obvious for the M1 radiative decay widths between the isoscalar $D_1N$ and $D_2^*N$ molecular pentaquarks, and their M1 radiative decay widths remain relatively stable when changing the binding energies for the initial and final molecular pentaquarks.
\item For the $[D_1N]({1}/{2}^+) \to [D_1N]({3}/{2}^+) \gamma$ and $[D_2^*N]({3}/{2}^+) \to [D_2^*N]({5}/{2}^+) \gamma$ processes, their M1 radiative decay widths are particularly small. This suppression is due to the limited phase spaces.
\item Compared to the M1 radiative decay widths between the isoscalar $DN$ and $D^*N$ molecular pentaquarks, the M1 radiative decay widths between the isoscalar $D_1N$ and $D_2^*N$ molecular pentaquarks are much smaller. {As shown in Eq.~(\ref{eq:M1}), the M1 radiative decay widths between hadrons depend on two key components: the phase space factor and the square of the amplitude. For the M1 radiative decay widths of single-charm molecular pentaquarks, the phase space factor is proportional to $k^3$, and we can derive
\begin{equation}
\left(\frac{k_{[D^*N ]\to [DN] \gamma}}{k_{[D_2^*N] \to [D_1N] \gamma}}\right)^3=74.
\end{equation}
Thus, the phase space factor of the $[D^*N ]\to [DN] \gamma$ process is considerably larger than that of the $[D_2^*N] \to [D_1N] \gamma$ process. In calculating the M1 radiative decay widths between the single-charm molecular pentaquarks, the square of the amplitude is proportional to the square of the transition magnetic moment $\left|\mu_{H \to H^{\prime}}\right|^2$. As illustrated in Fig. \ref{TMMCDHDT}, the absolute value of the transition magnetic moment of the $[D^*N ]\to [DN] \gamma$ process is comparable to that of the $[D_2^*N] \to [D_1N] \gamma$ process in most cases, except that the $[D_2^*N]({3}/{2}^+) \to [D_1N]({3}/{2}^+) \gamma$ process is slightly smaller. Based on the above discussions, it can be concluded that the suppression of the phase space is the primary factor contributing to the reduction in the M1 radiative decay width of the $[D_2^*N] \to [D_1N] \gamma$ process in comparison to the $[D^*N ]\to [DN] \gamma$ process.}
\end{enumerate}

\subsection{E1 radiative decays of the isoscalar $D_1N$ and $D_2^*N$ molecules to the isoscalar $DN$ and $D^*N$ molecules}

The above calculations show that the M1 radiative decay widths of the $D_1N$ and $D_2^*N$ states are very small, which is due to the close masses of the initial and final states, resulting in small photon momenta. It is evident that $D_1$ and $D_2^*$ are orbital excited charmed mesons, which have a significant mass gap with $D$ and $D^*$. Thus, the calculation of the $D^{(*)}\gamma$ transitions of the $D_1N$ and $D_2^*N$ molecules is a natural choice. In this type of transition, the orbital quantum numbers of the charmed mesons are changed. According to the properties of radiative decays, the E1 transitions may play a key role in these processes. Thus, we can consider the radiative decays of the $D_1N$ and $D_2^*N$ molecular states to the $DN$ and $D^*N$ molecular states via the E1 transitions.

\renewcommand\tabcolsep{0.12cm}
\renewcommand{\arraystretch}{1.50}
\begin{table}[!htbp]
\centering
\caption{The E1 radiative decay widths of the isoscalar $D_1N$ and $D_2^*N$ molecular pentaquarks to the isoscalar $DN$ and $D^*N$ molecular pentaquarks in units of keV. In view of the fact that these discussed molecular pentaquark candidates have not yet been definitively identified or observed in experiments, we employ the binding energies for the initial and final molecular pentaquark candidates in the range of $-2$ to $-14$ MeV to perform the quantitative calculations.}
\label{tab:E1D1ND2N}
\begin{tabular*}{86mm}{@{\extracolsep{\fill}}lcc}
\toprule[1.00pt]
\multicolumn{1}{c}{Radiative decays} & \makecell[c]{$S$-$D$ mixing\\ analysis}&\makecell[c]{Coupled channel\\ analysis}\\
\midrule[0.75pt]
$[D_1N](\frac{1}{2}^+)\to [D  N](\frac{1}{2}^-)  \gamma$&10.40$\sim$23.90 &9.50$\sim$20.60\\
$[D_1N](\frac{1}{2}^+)\to [D^*N](\frac{1}{2}^-)  \gamma$& 3.30$\sim$5.00  &3.30$\sim$5.10\\
$[D_1N](\frac{1}{2}^+)\to [D^*N](\frac{3}{2}^-)  \gamma$& 1.90$\sim$3.10  &1.90$\sim$3.20\\
$[D_1N](\frac{3}{2}^+)\to [D  N](\frac{1}{2}^-)  \gamma$&12.00$\sim$28.30 &11.10$\sim$23.90\\
$[D_1N](\frac{3}{2}^+)\to [D^*N](\frac{1}{2}^-)  \gamma$& 0.90$\sim$1.50  &0.90$\sim$1.50\\
$[D_1N](\frac{3}{2}^+)\to [D^*N](\frac{3}{2}^-)  \gamma$& 5.30$\sim$9.20  &5.30$\sim$9.20\\
$[D_2^*N](\frac{3}{2}^+)\to [D  N](\frac{1}{2}^-)\gamma$&0              &$\sim$0\\
$[D_2^*N](\frac{3}{2}^+)\to [D^*N](\frac{1}{2}^-)\gamma$&13.50$\sim$23.50 &13.50$\sim$23.50\\
$[D_2^*N](\frac{3}{2}^+)\to [D^*N](\frac{3}{2}^-)\gamma$&3.10$\sim$5.90   &3.10$\sim$5.90\\
$[D_2^*N](\frac{5}{2}^+)\to [D  N](\frac{1}{2}^-)\gamma$&0              &0\\
$[D_2^*N](\frac{5}{2}^+)\to [D^*N](\frac{1}{2}^-)\gamma$&0              &0\\
$[D_2^*N](\frac{5}{2}^+)\to [D^*N](\frac{3}{2}^-)\gamma$&21.70$\sim$40.90 &21.70$\sim$40.90\\
\bottomrule[1.00pt]
\end{tabular*}
\end{table}

The numerical results are presented in Table~\ref{tab:E1D1ND2N}. Here, we consider the $S$-$D$ mixing effect and the coupled channel effect. In most cases, the coupled channel effect leads to only small changes for these decay widths. According to Table~\ref{tab:E1D1ND2N}, we can obtain:
\begin{enumerate}[(i)]
\item For the $[D_1N]({1}/{2}^+)\to [D  N]({1}/{2}^-)\gamma$, $[D_1N]({3}/{2}^+)\to [D  N]({1}/{2}^-)\gamma$, $[D_2^*N]({3}/{2}^+)\to [D^*N]({1}/{2}^-)\gamma$, and $[D_2^*N]({5}/{2}^+)\to [D^*N]({3}/{2}^-)\gamma$, the E1 radiative decay widths are about 10 to 40 keV, which is several orders of magnitude larger than the results for the M1 radiative decay widths.
\item For $[D_1N]({1}/{2}^+)\to [D^*N]({1}/{2}^-)\gamma$, $[D_1N]({1}/{2}^+)\to [D^*N]({3}/{2}^-)\gamma$, $[D_1N]({3}/{2}^+)\to [D^*N]({1}/{2}^-)\gamma$, $[D_1N]({3}/{2}^+)\to [D^*N]({3}/{2}^-)\gamma$, and $[D_2^*N]({3}/{2}^+)\to [D^*N]({3}/{2}^-)\gamma$, the E1 radiative decay widths are estimated to be several keV. In these decay channels, the E1 radiative decay widths remain significantly larger than the M1 decays.
\item For the $[D_2^*N]({3}/{2}^+)\to [D  N]({1}/{2}^-)\gamma$ process, the E1 transition is forbidden in the $S$-$D$ mixing scheme. However, the $[D_2^*N]({3}/{2}^+)\to [D^*N]({1}/{2}^-)\gamma$ process is allowed for the E1 decay. In principle, if we consider the coupled channel effect of the $D N({1}/{2}^-)$-$D^* N({1}/{2}^-)$ system, it is possible that the $[D_2^* N]({3}/{2}^+)$ state can decay into $[D N]({1}/{2}^-) \gamma$ in the E1 mode with a narrow width. Nevertheless, the $D^*  N({1}/{2}^-)$ channel occupies only a very small component, and the obtained  width is still negligible. For the $[D_2^*N]({5}/{2}^+)\to [D  N]({1}/{2}^-)\gamma$ and $[D_2^*N]({5}/{2}^+)\to [D^*N]({1}/{2}^-)\gamma$, the E1 transition is forbidden for both the $S$-$D$ mixing effect and the coupled channel effect, since the M1 and E1 decays only work with $|J_H-J_{H^\prime}|\leq 1$.
\end{enumerate}

According to the above calculations, the E1 decay mode is very important for the $D_1N$ and $D_2^*N$ molecular states, and we propose to measure the $[D^{(*)}N]\gamma$ channels of the $D_1N$ and $D_2^*N$ molecular states in the future experiments.

In order to perform a comparative analysis of the E1 radiative decays between the $D_1$ and $D_2^*$ to the $D$ and $D^*$ as well as the isoscalar $D_1N$ and $D_2^*N$ molecules to the isoscalar $DN$ and $D^*N$ molecules, we calculate the E1 radiative decay widths of the $D_1$ and $D_2^*$ to the $D$ and $D^*$ by adopting the constituent quark model. In the following, we list the obtained numerical results:
\begin{eqnarray}
\renewcommand\tabcolsep{1.50cm}
\renewcommand{\arraystretch}{1.50}
\begin{array}{*{2}c}
\hline
\Gamma_{D_1^{+} \to D^{+} \gamma}\,(\mathrm{keV})~~~~~~~~~~~~& 25.32\nonumber\\
\Gamma_{D_1^{0} \to D^{0} \gamma}\,(\mathrm{keV})~~~~~~~~~~~~& 405.13\nonumber\\
\Gamma_{D_1^{+} \to D^{*+} \gamma}\,(\mathrm{keV})~~~~~~~~~~~~& 8.96\nonumber\\
\Gamma_{D_1^{0} \to D^{*0} \gamma}\,(\mathrm{keV})~~~~~~~~~~~~& 143.33\nonumber\\
\Gamma_{D_2^{*+} \to D^{+} \gamma}\,(\mathrm{keV})~~~~~~~~~~~~& 0\nonumber\\
\Gamma_{D_2^{*0} \to D^{0} \gamma}\,(\mathrm{keV})~~~~~~~~~~~~&0\nonumber\\
\Gamma_{D_2^{*+} \to D^{*+} \gamma}\,(\mathrm{keV})~~~~~~~~~~~~& 32.78\nonumber\\
\Gamma_{D_2^{*0} \to D^{*0} \gamma}\,(\mathrm{keV})~~~~~~~~~~~~& 524.47\nonumber\\
\hline
\end{array}.
\end{eqnarray}
Similar to the M1 radiative decays of the $[D^*N] \to [DN] \gamma$ and $D^* \to D \gamma$ processes, the spectator particle nucleon also plays a significant role in the E1 radiative decays of the isoscalar $D_1N$ and $D_2^*N$ molecules to the isoscalar $DN$ and $D^*N$ molecules.

\section{Magnetic moment properties}\label{sec3}

\subsection{Magnetic moment properties of the isoscalar $DN$, $D^*N$, $D_1N$, and $D_2^*N$ molecular pentaquarks}

As another important electromagnetic property unveiling the inner structures of the hadrons, in this subsection we further study the magnetic moment properties of the isoscalar $DN$, $D^*N$, $D_1N$, and $D_2^*N$ molecular pentaquarks. In our specific calculations, we consider the $S$-$D$ wave mixing and coupled channel effects. In Table \ref{MM}, we list the magnetic moment properties of the isoscalar $DN$, $D^*N$, $D_1N$, and $D_2^*N$ molecular pentaquarks.

\renewcommand\tabcolsep{0.25cm}
\renewcommand{\arraystretch}{1.50}
\begin{table}[!htbp]
\centering
\caption{The magnetic moment properties of the isoscalar $DN$, $D^*N$, $D_1N$, and $D_2^*N$ molecules. The units of the magnetic moments of the hadrons are $\mu_N$. Case I, Case II, and Case III correspond to the results obtained from the single channel analysis, the $S$-$D$ wave mixing analysis, and the coupled channel analysis, respectively. In light of the fact that these discussed molecular pentaquark candidates have not yet been definitively identified or observed in experiments, we employ three typical binding energies of $-2$, $-8$, and $-14~{\rm MeV}$ for these molecular pentaquark candidates in order to quantitatively calculate their magnetic moment properties.}\label{MM}
\begin{tabular}{cccc}\toprule[1pt]
Molecules&Case I&Case II&Case III\\\midrule[1.0pt]
$[DN](\frac{1}{2}^-)$&0.47&\multicolumn{1}{c}{/}&0.52,\,0.56,\,0.58\\
$[D^*N](\frac{3}{2}^-)$&0.37&0.37,\,0.37,\,0.37&\multicolumn{1}{c}{/}\\
$[D^*N](\frac{1}{2}^-)$&$-0.22$&$-0.16$,\,$-0.12$,\,$-0.09$&\multicolumn{1}{c}{/}\\
$[D_1N](\frac{3}{2}^+)$&$0.74$&$0.72$,\,$0.72$,\,$0.72$&$0.74$,\,$0.74$,\,$0.74$\\
$[D_1N](\frac{1}{2}^+)$&$0.03$&$0.04$,\,$0.05$,\,$0.06$&\multicolumn{1}{c}{/}\\
$[D_2^*N](\frac{5}{2}^+)$&$0.05$&$0.04$,\,$0.04$,\,$0.04$&\multicolumn{1}{c}{/}\\
$[D_2^*N](\frac{3}{2}^+)$&$-0.66$&$-0.62$,\,$-0.59$,\,$-0.57$&\multicolumn{1}{c}{/}\\
\bottomrule[1pt]
\end{tabular}
\end{table}

From Table \ref{MM}, the magnetic moment properties of the isoscalar $DN$, $D^*N$, $D_1N$, and $D_2^*N$ molecular pentaquarks are the important physical observables to unveil their inner structures, which can provide the critical suggestions for determining their spin-parity quantum numbers in the future experimental studies. For example, there exists the significant difference for the magnetic moment properties between the isoscalar $D^*N$ state with $J^P=1/2^-$ and the isoscalar $D^*N$ state with $J^P=3/2^-$, which depends on their inner structures.  By comparing the numerical results obtained from the single channel analysis, the $S$-$D$ wave mixing and coupled channel effects can affect the magnetic moment properties of the isoscalar $DN$, $D^*N$, $D_1N$, and $D_2^*N$ molecules, but the variation of their magnetic moments is less than $0.13\,{\mu_N}$.

\begin{figure}[htbp]
\centering
 \includegraphics[width=0.48\textwidth]{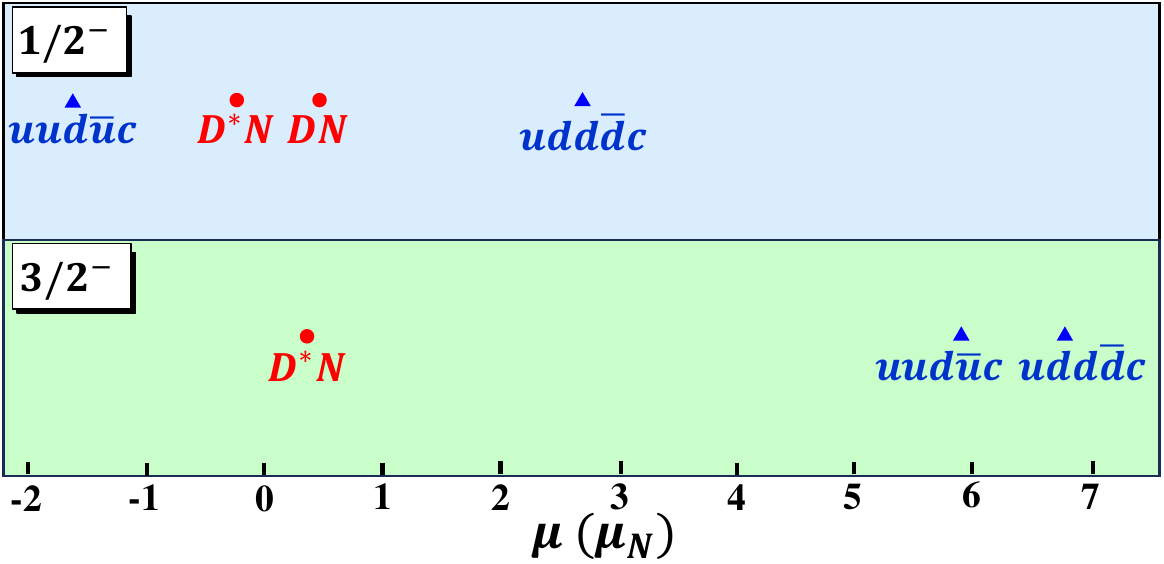}
\caption{The comparison focuses on the magnetic moments of the single-charm molecular and compact pentaquarks \cite{Sharma:2024wpc} with the identical quantum numbers and quark configurations.}\label{MMC}
\end{figure}

In Ref. \cite{Sharma:2024wpc}, the authors have performed a systematic calculation of the magnetic moment properties of the single-charm compact pentaquarks with $J^P=(1/2^-,\,3/2^-,\,5/2^-)$. Among them, the $uud\bar uc/udd\bar dc$ compact pentaquarks and the $DN/D^*N$ molecular pentaquarks may have the same quantum numbers and quark configurations. In Fig. \ref{MMC}, we compare the magnetic moment properties of the single-charm molecular and compact pentaquarks with the same quantum numbers and quark configurations. As shown in Fig. \ref{MMC}, there exists the significant differences for the magnetic moment properties of the single-charm molecular and compact pentaquarks with the identical quantum numbers and quark configurations. This is particularly evident in the comparison between the $D^*N$ molecular state with $I(J^P)=0(3/2^-)$ and the $udd\bar dc$ compact pentaquark with $I(J^P)=0(3/2^-)$. Thus, the magnetic moment properties can provide the key observables for distinguishing the single-charm molecular and compact pentaquarks with the same quantum numbers and quark configurations.

\subsection{Magnetic moments of the $\Lambda_c^{+} (2940)$ and $\Lambda_c^{+} (2910)$ under the molecular state and the charmed baryon interpretations}

In the following, we discuss whether the different theoretical explanations \cite{He:2006is,Garcia-Recio:2008rjt,Dong:2009tg,Dong:2010xv,He:2010zq,Liang:2011zza,Dong:2011ys,Ortega:2012cx,Zhang:2012jk,Ortega:2013fta,Dong:2014ksa,Ortega:2014eoa,Xie:2015zga,Yang:2015eoa,Zhao:2016zhf,Zhang:2019vqe,Wang:2020dhf,Yan:2022nxp,Xin:2023gkf,Ozdem:2023eyz,Yan:2023ttx,Yue:2024paz,Cheng:2006dk,Chen:2007xf,Ebert:2007nw,Zhong:2007gp,Liu:2009zg,Klempt:2009pi,Chen:2009tm,Cheng:2012fq,Lu:2016ctt,Lu:2018utx,Guo:2019ytq,Luo:2019qkm,Gandhi:2019xfw,Gong:2021jkb,Yu:2022ymb,Azizi:2022dpn,Zhang:2022pxc,Yu:2023bxn,Yang:2023fsc} of the $\Lambda_c^{+} (2940)$ and $\Lambda_c^{+} (2910)$ can be clarified by studying their magnetic moment properties.
Here, we present how to calculate the magnetic moment of the observed $\Lambda_c^{+} (2940)$ as the singly charmed baryon $\Lambda_c(2P,\,3/2^-)$. For the $\Lambda_c(2P,\,3/2^-)$, the overlaps of both color and spatial wave functions are 1. Therefore, it is necessary to construct the flavor and spin-orbital wave functions when calculating the magnetic moment of the $\Lambda_c(2P,\,3/2^-)$. For the $\Lambda_c(2P,\,3/2^-)$, the flavor wave function can be expressed as $\chi^{\rm flavor}=\frac{1}{\sqrt{2}}\left(ud-du\right)c$, while the spin-orbital wave function can be constructed by means of the following coupling
\begin{eqnarray}
[[[s_{q_1}s_{q_2}]_{s_{\ell}}[l_{\rho}l_{\lambda}]_L]_{j_{\ell}}s_c]_{JM}.
\end{eqnarray}
In this context, the symbols $s_{q_1}$, $s_{q_2}$, and $s_{c}$ represent the spins of the involved quarks. $s_{\ell}$ and $j_{\ell} $ are the spin and the total angular momentum of the light degree of freedom, respectively. $l_{\rho}$, $l_{\lambda}$, and $L$ are used to denote the $\rho$-mode, the $\lambda$-mode, and the total orbital angular momentum, respectively. Here, we use the $\rho$-mode to represent the coordinate between the two light flavor quarks, while we take the $\lambda$-mode to represent the relative position of the charmed quark and the center-of-mass of the two light flavor quarks. For the $\Lambda_c(2P,\,3/2^-)$, we can obtain $s_{q_1}=s_{q_2}=s_c=1/2$, $s_{\ell}=0$, $l_{\rho}=0$, $l_{\lambda}=1$, $L=1$, and $j_{\ell}=1$. Thus, the spin-orbital wave function of the $\Lambda_c(2P,\,3/2^-)$ with $(J,\,M)=(3/2,\,3/2)$ can be explicitly expressed as
\begin{eqnarray}
\chi^{\rm spin-orbital}=&-&\sqrt{\frac{1}{2}}\left|\frac{1}{2},-\frac{1}{2}\right\rangle^{s_{q_1}}\left|\frac{1}{2},\frac{1}{2}\right\rangle^{s_{q_2}}\left|\frac{1}{2},\frac{1}{2}\right\rangle^{s_c} Y_{0,0}^{l_{\rho}}Y_{1,1}^{l_{\lambda}}\nonumber\\
&+&\sqrt{\frac{1}{2}}\left|\frac{1}{2},\frac{1}{2}\right\rangle^{s_{q_1}}\left|\frac{1}{2},-\frac{1}{2}\right\rangle^{s_{q_2}}\left|\frac{1}{2},\frac{1}{2}\right\rangle^{s_c} Y_{0,0}^{l_{\rho}}Y_{1,1}^{l_{\lambda}}.\nonumber\\
\end{eqnarray}
According to the above preparation, the magnetic moment of the singly charmed baryon $\Lambda_c(2P,\,3/2^-)$ can be calculated by the following matrix element
\begin{eqnarray}
&&\left\langle{\chi^{\rm flavor}\otimes\chi^{\rm spin-orbital}\left|\sum_{j}\hat{\mu}_{zj}^{\rm spin}+\hat{\mu}_z^{\rm orbital}\right|\chi^{\rm flavor}\otimes\chi^{\rm spin-orbital}}\right\rangle\nonumber\\
&&=\mu_c+\mu_{[ud]c}^L.
\end{eqnarray}
Here, $\mu_{[ud]c}^L=\frac{m_{[ud]}}{m_{[ud]}+m_{c}}\frac{e_c}{2m_c}+\frac{m_{c}}{m_{[ud]}+m_{c}}\frac{e_{[ud]}}{2m_{[ud]}}$, while $e_{[ud]}$ and $m_{[ud]}$ are the charge and the mass of the diquark $[ud]$, respectively.

\begin{figure}[htbp]
\centering
\includegraphics[width=0.30\textwidth]{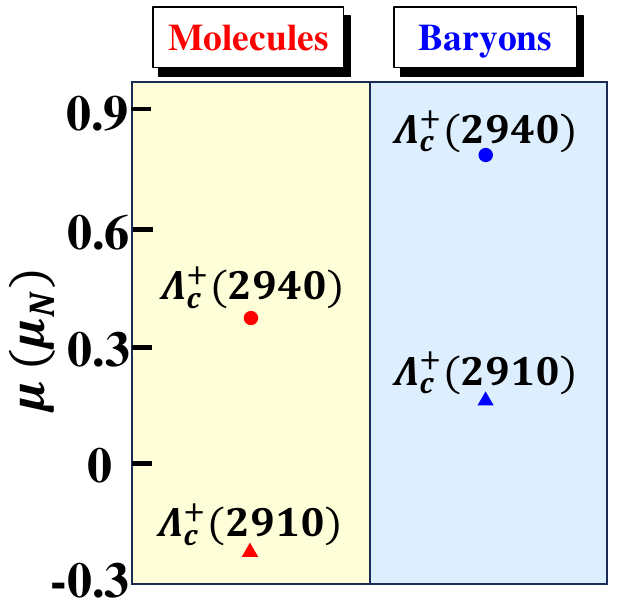}
\caption{The magnetic moment properties of the $\Lambda_c^{+} (2940)$ and $\Lambda_c^{+} (2910)$ within the frameworks of the $D^*N$ molecular states and the $2P$ states of singly charmed baryon, respectively. {For the magnetic moments of the $\Lambda_c^{+} (2940)$ and $\Lambda_c^{+} (2910)$ under the $D^*N$ molecular states, we only list the results of the single channel analyses.}}\label{M29402910}
\end{figure}

In Fig. \ref{M29402910}, we present an analysis of the magnetic moment properties of the $\Lambda_c^{+} (2940)$ and $\Lambda_c^{+} (2910)$ within the frameworks of the $D^*N$ molecular states and the $2P$ states of singly charmed baryon, respectively. If the $\Lambda_c^{+} (2940)$ and $\Lambda_c^{+} (2910)$ can be interpreted as the $D^*N$ molecular states with $I(J^P)=0(3/2^-)$ and $0(1/2^-)$ \cite{He:2006is,Garcia-Recio:2008rjt,Dong:2009tg,Dong:2010xv,He:2010zq,Liang:2011zza,Dong:2011ys,Ortega:2012cx,Zhang:2012jk,Ortega:2013fta,Dong:2014ksa,Ortega:2014eoa,Xie:2015zga,Yang:2015eoa,Zhao:2016zhf,Zhang:2019vqe,Wang:2020dhf,Yan:2022nxp,Xin:2023gkf,Ozdem:2023eyz,Yan:2023ttx,Yue:2024paz}, the magnetic moments of the $\Lambda_c^{+} (2940)$ and $\Lambda_c^{+} (2910)$ are around $0.37\,\mu_{N}$ and $-0.22\,\mu_{N}$, respectively. If the $\Lambda_c^{+} (2940)$ and $\Lambda_c^{+} (2910)$ can be explained as the singly charmed baryons $\Lambda_c(2P,\,3/2^-)$ and $\Lambda_c(2P,\,1/2^-)$ \cite{Cheng:2006dk,Chen:2007xf,Ebert:2007nw,Zhong:2007gp,Liu:2009zg,Klempt:2009pi,Chen:2009tm,Cheng:2012fq,Lu:2016ctt,Lu:2018utx,Guo:2019ytq,Luo:2019qkm,Gandhi:2019xfw,Gong:2021jkb,Yu:2022ymb,Azizi:2022dpn,Zhang:2022pxc,Yu:2023bxn,Yang:2023fsc}, the magnetic moments of the $\Lambda_c^{+} (2940)$ and $\Lambda_c^{+} (2910)$ are approximately $0.79\,\mu_{N}$ and $0.16\,\mu_{N}$, respectively. As illustrated in Fig. \ref{M29402910}, there exist the difference for the magnetic moment properties of the $\Lambda_c^{+} (2940)$ and $\Lambda_c^{+} (2910)$ within the frameworks of the $D^*N$ molecular states and the $2P$ states of singly charmed baryon. Thus, the different theoretical explanations for the $\Lambda_c^{+} (2940)$ and $\Lambda_c^{+} (2910)$ can be clarified by studying the magnetic moment properties in the future experiments. We hope further experiments can bring us more surprises.

\section{Discussions and conslusions}\label{sec4}

Hadron spectroscopy is a crucial tool for unraveling the non-perturbative dynamics of the strong interaction. Exotic states, as important members of the hadron family, have garnered considerable attention. A notable example is the observation of the $P_{\psi}^{N}(4312)$, $P_{\psi}^{N}(4440)$, and $P_{\psi}^{N}(4457)$ states \cite{Aaij:2019vzc}, which strongly supports the existence of the hidden-charm molecular pentaquark states \cite{Li:2014gra,Karliner:2015ina,Wu:2010jy,Wang:2011rga,Yang:2011wz,Wu:2012md,Chen:2015loa}. Among various molecular-type pentaquarks, the single-charm molecular pentaquarks composed of charmed mesons and nucleons have been studied by presenting their mass spectrum. However, this is not the whole story of their spectroscopic behavior. Additionally, the $\Lambda_c^{+}(2940)$ and $\Lambda_c^{+}(2910)$ \cite{BaBar:2006itc,Belle:2006xni,LHCb:2017jym,Belle:2022hnm} can be considered as single-charm molecular pentaquark candidates \cite{He:2006is,Garcia-Recio:2008rjt,Dong:2009tg,Dong:2010xv,He:2010zq,Liang:2011zza,Dong:2011ys,Ortega:2012cx,Zhang:2012jk,Ortega:2013fta,Dong:2014ksa,Ortega:2014eoa,Xie:2015zga,Yang:2015eoa,Zhao:2016zhf,Zhang:2019vqe,Wang:2020dhf,Yan:2022nxp,Xin:2023gkf,Ozdem:2023eyz,Yan:2023ttx,Yue:2024paz}, though other conventional explanations exist \cite{Cheng:2006dk,Chen:2007xf,Ebert:2007nw,Zhong:2007gp,Liu:2009zg,Klempt:2009pi,Chen:2009tm,Cheng:2012fq,Lu:2016ctt,Lu:2018utx,Guo:2019ytq,Luo:2019qkm,Gandhi:2019xfw,Gong:2021jkb,Yu:2022ymb,Azizi:2022dpn,Zhang:2022pxc,Yu:2023bxn,Yang:2023fsc}. Distinguishing between different assignments of the $\Lambda_c^{+}(2940)$ and $\Lambda_c^{+}(2910)$ remains a challenging task.

In the present work, we carry out a systematic investigation of the electromagnetic properties of the single-charm molecular pentaquark candidates, and discuss the electromagnetic properties of the observed $\Lambda_c^{+} (2940)$ and $\Lambda_c^{+} (2910)$ in the $2P$ states of singly charmed baryon. In our concrete calculations, we adopt the constituent quark model, while both the $S$-$D$ wave mixing effect and the coupled channel effect are taken into account. Firstly, we study the M1 radiative decay widths of the isoscalar $DN$, $D^*N$, $D_1N$, and $D_2^*N$ molecular pentaquarks. Our numerical results show that the M1 radiative decay widths between the isoscalar $DN$, $D^*N$, $D_1N$, and $D_2^*N$ molecular pentaquarks provide some insights into their inner structures, which can serve as the crucial clues for distinguishing their spin-parity quantum numbers in the future experimental studies, and the spin-parity quantum numbers of the $\Lambda_c^{+} (2940)$ and $\Lambda_c^{+} (2910)$ in the framework of the $D^*N$ molecular states can be clarified by studying their M1 radiative decay widths. Besides, we also discuss the E1 radiative decay widths for the $D_1N$ and $D_2^*N$ molecules to the $DN$ and $D^*N$ molecules, which is very important for the $D_1N$ and $D_2^*N$ molecular states.

After that, we study the magnetic moment properties of the isoscalar $DN$, $D^*N$, $D_1N$, and $D_2^*N$ molecular pentaquarks. The magnetic moment properties of the isoscalar $DN$, $D^*N$, $D_1N$, and $D_2^*N$ molecular pentaquarks can provide the important observables to decipher their inner structures, which can serve as the crucial clues for distinguishing their spin-parity quantum numbers and configurations in experimental studies. In particular, the magnetic moment properties of the $\Lambda_c^{+} (2940)$ and $\Lambda_c^{+} (2910)$ are different within the frameworks of the $D^*N$ molecules and the $2P$ states of singly charmed baryon, and the magnetic moment properties of the single-charm molecular and compact pentaquarks with the identical quantum numbers and quark configurations exhibit the significant differences.

Since 2015, the significant progress has been made in the study of the hidden-charm molecular pentaquarks. As an interesting and important research topic, the search for the single-charm molecular pentaquarks will be a new task for the exploration of the pentaquark states in the future experiments. Thus, more theoretical efforts should be made to provide the abundant suggestions for the search of the single-charm molecular pentaquarks. Obviously, the study of their electromagnetic characteristics can provide the valuable information for the future experimental construction of the family of the single-charm molecular pentaquarks.

\section*{Acknowledgement}

This work is supported by the National Natural Science Foundation of China under Grant Nos. 12405097, 12335001, 12247155, 12247101 and 12405098, the China National Funds for Distinguished Young Scientists under Grant No. 11825503, National Key Research and Development Program of China under Contract No. 2020YFA0406400, the 111 Project under Grant No. B20063, the fundamental Research Funds for the Central Universities, and the project for top-notch innovative talents of Gansu province. F.L.W. is also supported by the China Postdoctoral Science Foundation under Grant No. 2022M721440.

\end{document}